\documentclass[journal]{IEEEtran}
\ifCLASSINFOpdf
\else
\fi

\usepackage{cite}
\usepackage{graphicx}
\usepackage{multicol}
\usepackage{multirow}
\usepackage{subcaption}
\usepackage{amsmath}
\usepackage{amsfonts}
\usepackage{romannum}
\usepackage{caption}
\usepackage{tabularx}
\usepackage{tabulary}
\usepackage{array}
\usepackage{tabu}
\usepackage{hhline}
\usepackage{makecell}
\usepackage{float}
\usepackage{stfloats}

\begin{document}

\title{Cardiac Segmentation on CT Images through \\Shape-Aware Contour Attentions}

\author{Sanguk Park and Minyoung Chung$^{\ast}$%
\thanks{\textit{Asterisk indicates corresponding author (Minyoung Chung).}}%
\thanks{S. Park is with the Lunit Inc., South Korea (e-mail: tony.superb@lunit.io).}%
\thanks{*M. Chung is with the School of Software, Soongsil University, South Korea (e-mail: chungmy@ssu.ac.kr).}%
}

\markboth{}%
{Shell \MakeLowercase{\textit{et al.}}: Bare Demo of IEEEtran.cls for IEEE Journals}

\maketitle

\begin{abstract}
Cardiac segmentation of atriums, ventricles, and myocardium in computed tomography (CT) images is an important first-line task for presymptomatic cardiovascular disease diagnosis. In several recent studies, deep learning models have shown significant breakthroughs in medical image segmentation tasks. Unlike other organs such as the lungs and liver, the cardiac organ consists of multiple substructures, i.e., ventricles, atriums, aortas, arteries, veins, and myocardium. These cardiac substructures are proximate to each other and have indiscernible boundaries (i.e., homogeneous intensity values), making it difficult for the segmentation network focus on the boundaries between the substructures. In this paper, to improve the segmentation accuracy between proximate organs, we introduce a novel model to exploit shape and boundary-aware features. We primarily propose a shape-aware attention module, that exploits distance regression, which can guide the model to focus on the edges between substructures so that it can outperform the conventional contour-based attention method. In the experiments, we used the Multi-Modality Whole Heart Segmentation dataset that has 20 CT cardiac images for training and validation, and 40 CT cardiac images for testing. The experimental results show that the proposed network produces more accurate results than state-of-the-art networks by improving the Dice similarity coefficient score by 4.97\%. \par
Our proposed shape-aware contour attention mechanism demonstrates that distance transformation and boundary features improve the actual attention map to strengthen the responses in the boundary area. Moreover, our proposed method significantly reduces the false-positive responses of the final output, resulting in accurate segmentation.
\end{abstract}


\IEEEpeerreviewmaketitle

\section{Introduction}
% \IEEEPARstart{C}{ardiac} ...\par
The heart is one of the most essential organs in the human body. It is often called the engine of the human body because it is located at the center of the chest and it supplies blood flow throughout the body. Therefore, even small abnormalities in the cardiac system can lead to fatal consequences. Cardiovascular diseases (CVDs) are one of the major causes of death worldwide \cite{McNamara2019}. On average, death due to CVDs occur once every 37s, which this indicates that 2,353 deaths occur each day in the United States \cite{doi:10.1161/CIR.0000000000000757}. According to the World Health Organization, 17.9 million people die each year from CVDs. It has been reported that up to 90\% of CVDs can be prevented \cite{doi:10.1161/CIRCULATIONAHA.107.717033, ODONNELL2016761} by in-time diagnosis of heart problems. Applications of computer-aided prediagnosis systems are rapidly developing owing to the advances in computerized imaging devices (e.g., computed tomography (CT) or magnetic resonance imaging (MRI)). Automated segmentation of cardiac CT images can aid in preventing CVDs, such as coronary artery disease, stroke, and cardiomyopathy, in presymptomatic subjects.\par

\begin{figure}[t]
    \centering
    \includegraphics[width=\linewidth]{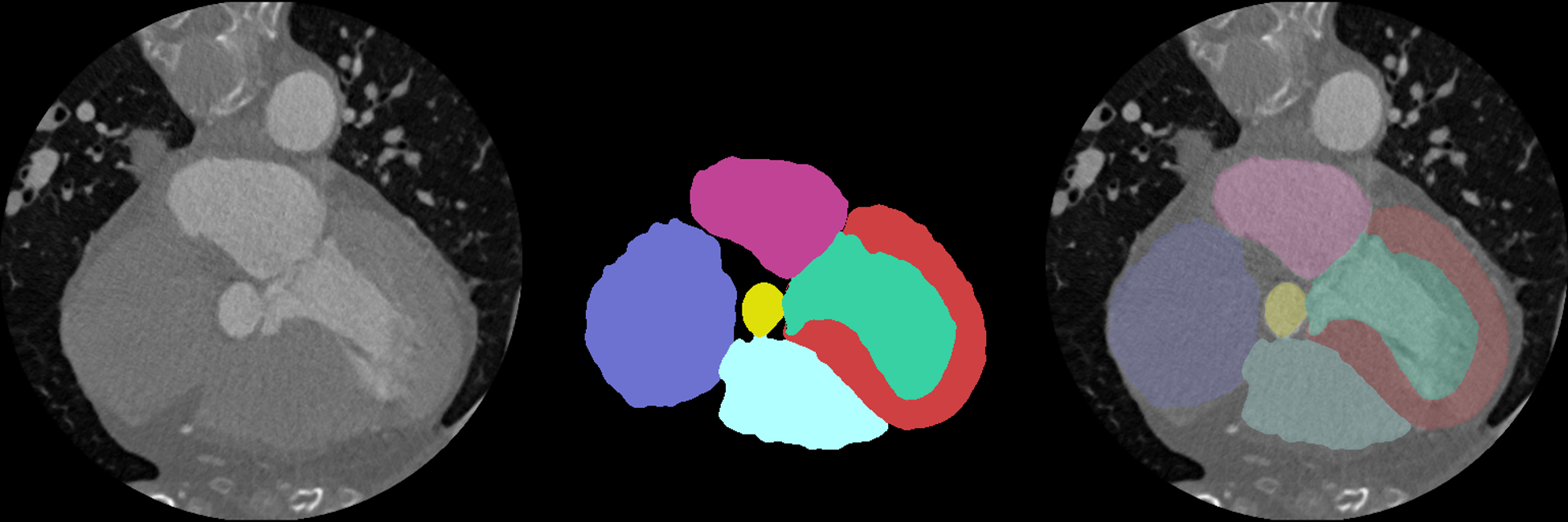}    \caption[Example data]{Cardiac computed tomography image, manual segmentation labels of cardiac substructures, and overlapped images}
    \label{fig:Intro_heart}
\end{figure}

% \begin{figure}[t]
%     \centering
%     \includegraphics[width=\linewidth]{figures/Intro_cardiacct.png}
%     \caption[Cardiac CT image]{Shape variations of cardiac CT images}
%     \label{fig:Intro_cardiacct}
% \end{figure}

In the past decade, deep learning in computer vision especially convolutional neural networks (CNNs) has achieved superior performance. CNNs have replaced several conventional tasks in the computer vision field, including image classification \cite{simonyan2015deep, he2015deep, Huang2017}, localization \cite{ren2016faster, redmon2018yolov3} and segmentation \cite{long2015fully}. While it is difficult to gather and handle large-scale medical datasets, CNNs have been widely employed for a vast number of automated medical image segmentation tasks \cite{ronneberger2015unet, milletari2016vnet, Gibson2018 ,oktay2018attention, chung2020liver}. Notably, \textit{Ronneberger et al.} \cite{ronneberger2015unet} successfully boosted the integration of CNNs into medical imaging. Their network, named U-Net \cite{ronneberger2015unet}, combined the skip connection and the encoder-decoder architecture to propagate low-level spatial information. In addition to the architectural advances, Dice similarity coefficient (DSC) loss was proposed by \textit{Milletari et al.} \cite{milletari2016vnet}, which increased the deep learning performance by resolving the data imbalance problem between foreground and background voxels. Particularly, several deep learning algorithms that are based on CNN have been proposed for cardiac imaging \cite{Li2017, Yu2017, Zheng2018, KHENED201921, Du2019}. A 2.5D multislice network \cite{Zheng2018, Du2019} was 
utilized for cardiac ventricle segmentation; however, the 3D spatial features were not fully exploited in the 2.5D methods, which employed multiple 2D sliced images. Automatic whole heart (WH) segmentation methods \cite{Li2017, Yu2017} have been researched based on conventional segmentation networks (e.g., FCN and U-Net). The results of the conventional methods showed inferior performances because these methods ignored the identification of accurate boundaries between cardiac substructures. The prior methods were also limited in terms of distinguishing between homogeneous intensity boundaries because of the difficulty in preserving both spatial and shape features.\par

\begin{figure}[t]
    \centering
    \includegraphics[width=\linewidth]{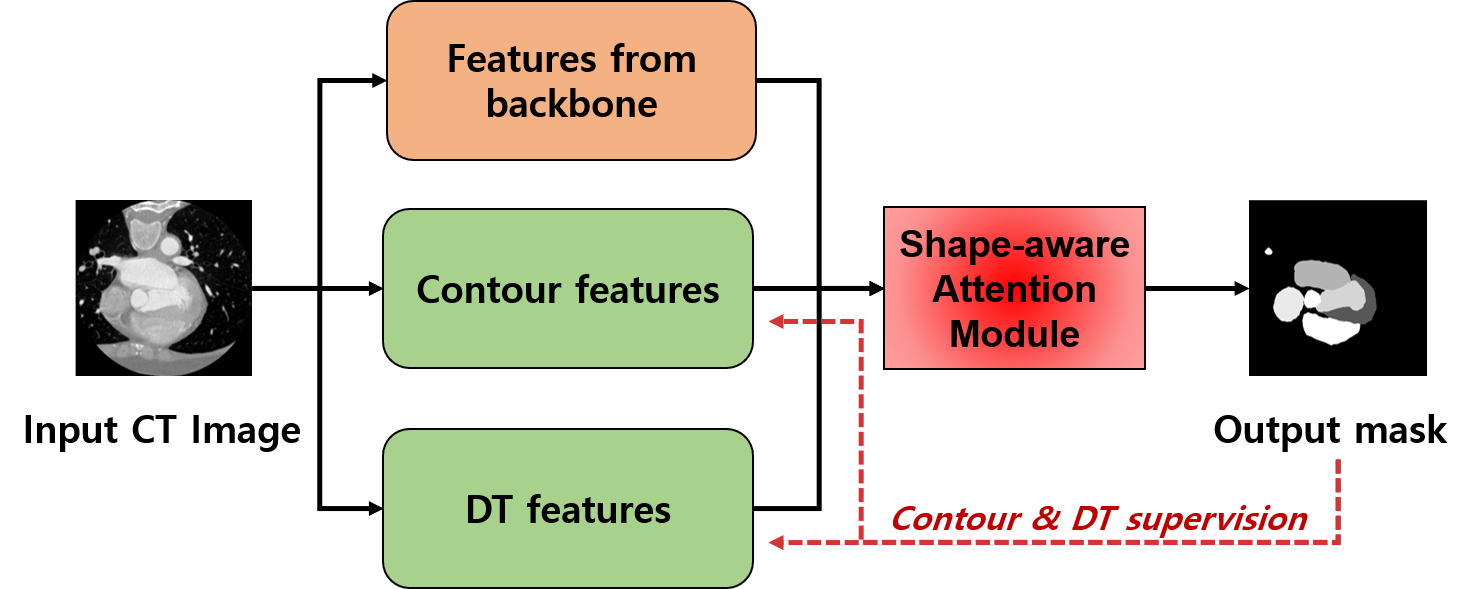}
    \caption{Overall architecture of the proposed convolutional neural network model\protect\footnotemark }
    \label{fig:overall}
\end{figure}

In this study, we aimed to improve the performance of cardiac segmentation in CT images. Accurate cardiac segmentation is a challenging problem because the heart is composed of several substructures. Fig. \ref{fig:Intro_heart} shows that the human heart is a complex organ consisting of four chambers, blood vessels, and muscles. The aim of this study was to precisely segment substructures, including the four chambers (left ventricle (LV), right ventricle (RV), left atrium (LA), right atrium (RA)), two arteries (ascending aorta (AA), pulmonary artery (PA)), and myocardium of the left ventricle (LV-myo). The major difficulty in segmenting cardiac suborgans is identifiying the boundary between adjacent organs (e.g., boundary between the atrium and the ventricle). To address this problem, we propose a CNN model with a novel attention mechanism (Fig. \ref{fig:overall}). To focus on the shape and boundary of cardiac structures, we conjugated the features of the distance transformation (DT) of the labeled ground-truth image, and the contour of the object into the attention mechanism. Specifically, the method drives the CNN model to better learn boundary-aware features based on shape-aware features (i.e., DT). Moreover, the proposed attention mechanism can produce exact segmentation results by reducing the false-positive responses. We employed our proposed shape-aware contour attention mechanism on a conventional encoder-decoder architecture U-Net \cite{ronneberger2015unet}.

\footnotetext{A DT represents a distance transformation}

The remainder of this patper is organized as follows. In section {\Romannum{2}}, the background, including the visual attention mechanism, shape-prior embedding, and DT are reviewed. The proposed network architecture and its details are described in section {\Romannum{3}}. The experimental results when compared to other CNNs and the ablations of our proposed network are presented in section {\Romannum{4}}. Finally, the discussion and conclusions are presented in section {\Romannum{5}}.

\section{Background}

\subsection{Visual Attention Mechanism}
The attention mechanism for CNN is widely used for image classification, localization and segmentation \cite{ba2015multiple, zhou2015learning, chen2016attention, wang2017residual}. The intention of the visual attention mechanism is simply to enhance the output of the receptive field around the target objects. \textit{Xu et al.} \cite{pmlr-v37-xuc15} showed how a model automatically focuses on objects in the image through visualization. \textit{Wang et al.} \cite{wang2017residual} stacked the attention modules for image classification. The Squeeze and Excitation block\cite{hu2019squeezeandexcitation}, a type of attention vector, was proposed to recalibrate the feature maps. Based on \cite{wang2017residual, hu2019squeezeandexcitation}, the bottleneck attention module (BAM) \cite{bam} and the convolutional block attention module (CBAM) \cite{cbam} were proposed. BAM and CBAM could be easily combined with the universal CNN models and achieved better performance. The channel attention module weighs the importance of features at the channel level, and the spatial attention module encodes the spatial location of objects in the feature maps. However, BAM and CBAM are limited while extracting shape-prior features because the max and average pooling functions weaken the details of shape features by reinforcing spatial location information. To force the model to refine the detailed segmentation of object boundaries, \textit{Zhuge et al.} \cite{8489930} employed contour-supervised features in the attention module. \textit{Chung et al.} \cite{chung2020liver} introduced a self-supervising contour attention to automatically identify a deep context for segmentation. However, by employing a single contour feature, it is difficult to obtain the attention map in a homogeneous intensity area. In this study, to complement the contour attention, both the DT and contour features are fed to the shape-aware contour attention mechanism to reduce the responses on the background area.\par

\subsection{Shape-prior Embedding}
In the biomedical image segmentation field, the visual attention mechanism was studied to extract shape-prior information from coarse features, based on the fact that most human organs, such as the liver and lung, typically form certain shapes. Based on this idea, DenseVNet was introduced by \textit{Gibson et al.} \cite{Gibson2018} to segment multiple organs. They attached auxiliary $12^3$ learnable parameters to capture the approximate shape of the target organs. The auxiliary shape-prior feature was concatenated to the final output, providing a hint to the model as an additive attention mechanism. \textit{Oktay et al.} \cite{oktay2018attention} conducted a self-attention mechanism focusing on shape-prior features. Based on the U-Net architecture, a soft-attention gating module was proposed to utilize spatial contextual information \cite{oktay2018attention}. The authors implemented a self-attention mechanism by connecting high-level features that are sufficient to express the shape-prior context as gating to the attention module. These networks outperformed the other networks by using spatial information; however, a trainable volume of $12^3$ parameters is too coarse to express the details of the shape. Moreover, the proposed self-attention mechanism cannot be applied for delineating the boundaries between the shapes of multiple organs.

\subsection{Distance Transformation for Convolutional Neural Network-based Image Segmentation}
The DT operation represents distances for each pixel (or voxel) in terms of a given binary image \cite{BORGEFORS1986344}. Typically, we use the boundary of the foreground area as a set of pixels, and in the DT process, pixels inside the foreground area have a distance to the nearest boundary. DT can be denoted as follows:
\begin{equation}
    DT(P)[a]=min_{y\in P} d(a,b)
    \label{eq:dt}
\end{equation}
where $P$ is a predetermined point set, such as the boundary of the foreground, or background, and $a, b$ are points. $d(a,b)$ indicates the function that calculates the distance between $a$ and $b$. \par
Because DT can represent shape information, it has been employed for image segmentation tasks. For example, the DT and watershed algorithms were combined to segment the objects of interest. In recent years, DT has been employed in CNNs, and has shown promising results. \textit{Audebert et al.} \cite{audebert2019distance} used signed DT as a regression target to detect clear boundaries and shapes. \textit{Wang et al.} \cite{doi:10.1146/annurev-bioeng-071516-044442} combined deep learning and the watershed algorithm to detect cells. \textit{Chung et al.} \cite{CHUNG2020103720} calculated the loss between tooth segmentation results and their ground-truth DT to learn similar tooth shape features. \textit{Navarro et al.} \cite{10.1007/978-3-030-32692-0_71} trained the model using labels, DT from labels, and the contour of the labeled images. For cardiac segmentation, \textit{Dangi et al.} \cite{Dangi2019} added a complementary decoder to directly regress the DT, and subsequently, the DT was used to regularize the network. The overall limitation of previous works is that they employed an auxiliary task of regressing DT; however, the shape-related features were not successfully trained by the proposed methods. More importantly, DT features have not been explicitly used to improve of boundary delineations.
\par

\begin{figure*}[ht!]
    \centering
    \includegraphics[width=\linewidth]{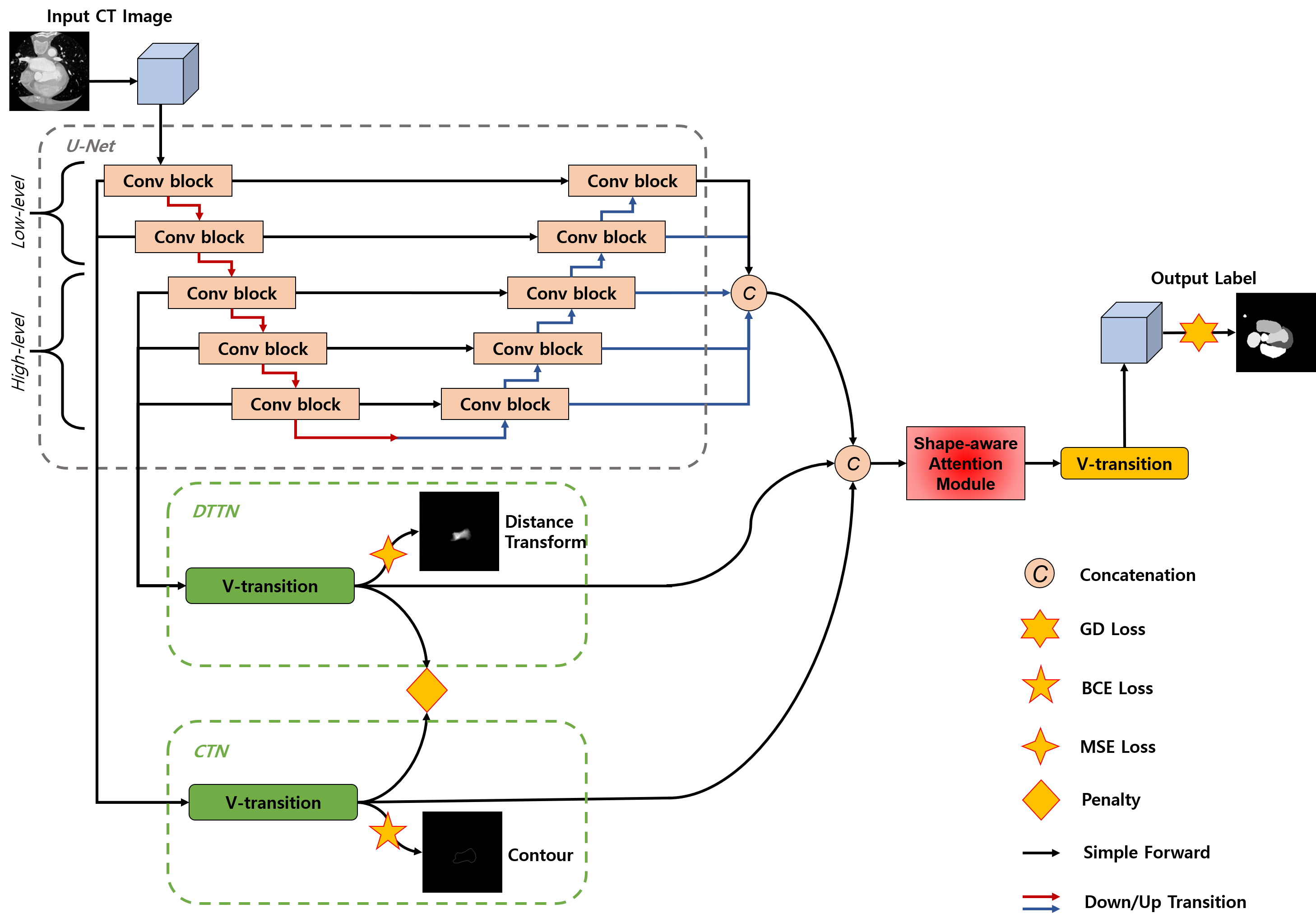}
    \caption{Schematic visualization of \textit{contour- and distance-transform-guided attention network (CDA-Net)}. The blue and yellow cubes correspond to feature maps. The dotted line denotes each feature map from the contour transition network and distance transform transition network.}
    \label{fig:Our_arch}
\end{figure*}

\begin{figure}[ht!]
    \centering
    \includegraphics[width=\linewidth]{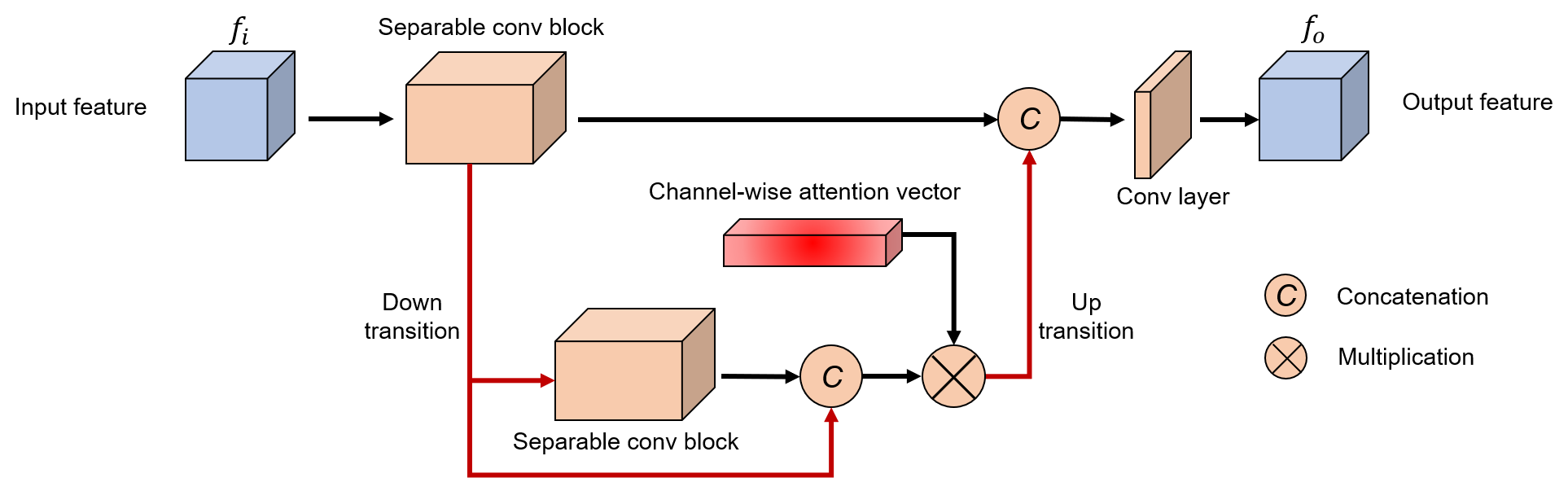}
    \caption{Diagram of V-transition module introduced in \cite{chung2020liver}, which is composed of a single-level encoder-decoder architecture and channel-wise attention. The separable conv block consists of a convolutional layer with groups, batch normalization, and a non-linear activation function, which reduces the computation times and parameters. The single-level encoder-decoder architecture supports the parameters to be trained with multi-scale spatial features, and the channel-wise attention refines and compresses the intermediate features.}
    \label{fig:Our_v}
\end{figure}

\section{Methodology}
\subsection{Contour- and Distance-transform-guided Attention Network Architecture}

To segment the entire heart from cardiac CT images using a deep learning technique, we designed a novel CNN named \textit{contour- and distance-transform-guided attention network} (CDA-Net). The proposed architecture is illustrated in Fig. \ref{fig:Our_arch}. The U-Net \cite{3dunet2016} architecture is used as the backbone of our model. Using the coarse features from the backbone, auxiliary V-transition \cite{chung2020liver} modules obtain the contour and DT features. In addition, a novel penalty energy is developed to mutually complement the contour and DT features. We further integrated these features (i.e., contour and DT) and the outputs of the backbone to feed the attention mechanism. In the following shape-aware attention module, the DT features suppress the feature responses in the background area, allowing the model to suppress false-positive responses. In contrast, the contour features force the model to focus on the detail of the object boundary. Finally, another V-transition is applied to aggregate and refine the features for multi-class segmentation results. The details of the proposed components are described in the subsequent paragraphs. \par

\vspace{2mm}
\noindent \textit{\textbf{Contour and Distance Transform Transitions}}
Based on the backbone network, the contour transition network (CTN) and distance transform transition network (DTTN) were designed to form CDA-Net. The underlying design principles for CTN and DTTN are to predict the contour probability of each voxel and coarse shape-prior information, respectively. The role of CTN and DTTN is to refine the following shape-aware attention module, which is developed to better represent the features of cardiac structures. From the encoder part of the backbone, low-level features are the fed to CTN to generate contour probabilistic map features, and high-level features are input to the DTTN to directly regress the DT features. We utilized the low-level features for the CTN because the contour is a combination of local edge features; conversely, high-level features were used to estimate DT, which requires global shape information for accurate prediction. As shown in Fig. \ref{fig:Our_arch}, each transition block represents a V-transition block (Fig. \ref{fig:Our_v}), which is a single-level encoder-decoder module with a channel-wise attention mechanism. The V-transition has advantages in learning multiscale features with only a few parameters, which can help to learn the shape of the heart from multiscale CT images \cite{chung2020liver}. For the multiclass task (i.e., multiple suborgans), class-wise contours and DT were individually used to train the CTN and DTTN, respectively. We applied binary cross-entropy (BCE) loss and mean squared error (MSE) loss to obtain the outputs of CTN and DTTN, respectively. The model predicts the contour probability per voxel for each class and computes the differences between the forwarded contour feature and the ground-truth using the following equation:
\begin{equation}
    L_{BCE}=-(ylog(p)+(l-y)log(l-p)).
    \label{eq:loss_bce}
\end{equation}
Furthermore, because there are only a few contour voxels when compared to the background voxels, we set the weights as [0.001, 0.999] to solve the highly imbalanced classification task. \par 
Similarly, to train the DT feature forthe  DTTN, the MSE loss is employed. The MSE loss for the predicted DT map $P$ and ground-truth $Y$ with $n$ number of voxels is defined as
\begin{equation}
    L_{MSE}=\cfrac{1}{n}\sum_{i=1}^{n}{(y_i - p_i)}^2.
    \label{eq:loss_mse}
\end{equation}
In summary, two V-transitions are employed to obtain the contour and DT features from the backbone. These additional features can help the model regress the details of object boundaries by feeding it to the attention mechanism.\par

\begin{figure}
    \centering
    \subcaptionbox{DT feature}[0.3\linewidth]{\includegraphics[width=\linewidth]{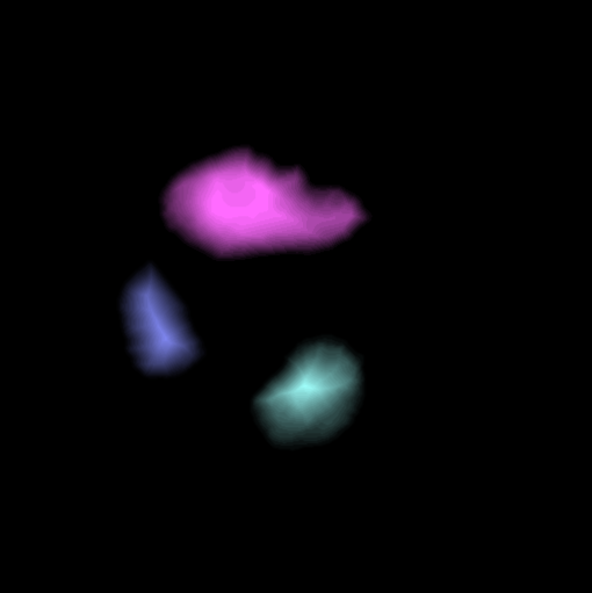}}
    \subcaptionbox{Inversion of clamped DT feature}[0.3\linewidth]{\includegraphics[width=\linewidth]{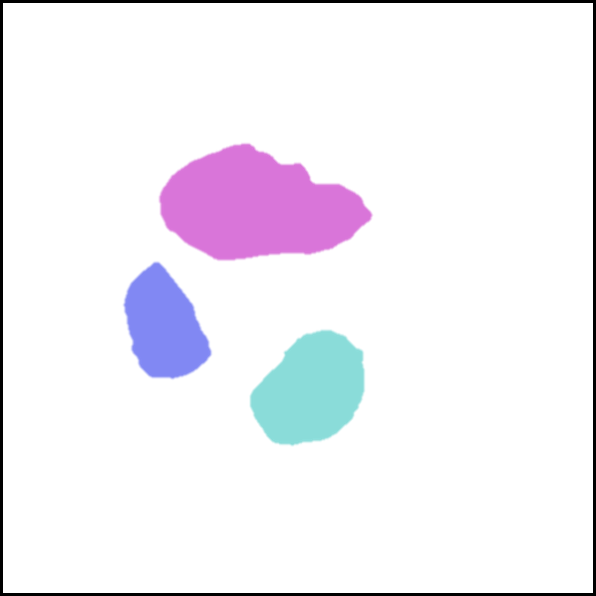}}
    \subcaptionbox{Contour feature}[0.3\linewidth]{\includegraphics[width=\linewidth]{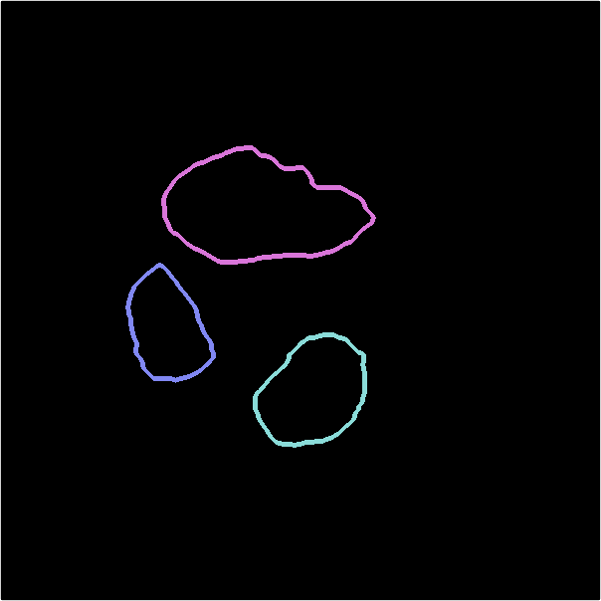}}
    
    \begin{subfigure}[]{0.3\linewidth}
        \includegraphics[width=\linewidth]{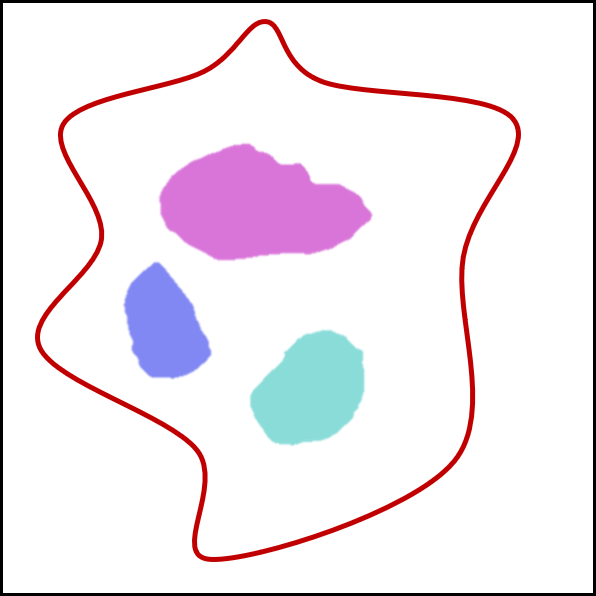}
        \subcaption{High penalty}
    \end{subfigure}
    \begin{subfigure}[]{0.3\linewidth}
        \includegraphics[width=\linewidth]{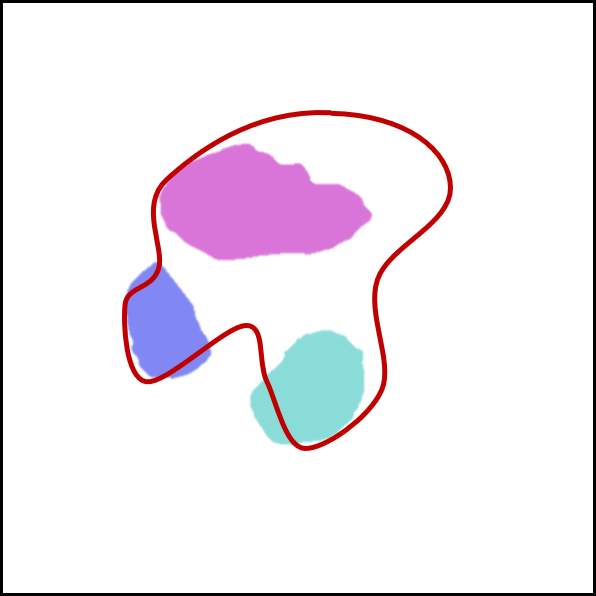}
        \subcaption{Middle penalty}
    \end{subfigure}
    \begin{subfigure}[]{0.3\linewidth}
        \includegraphics[width=\linewidth]{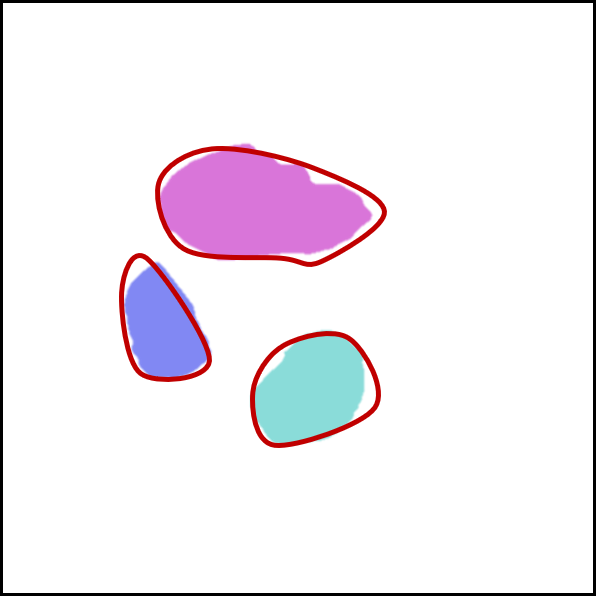}
        \subcaption{Low penalty}
    \end{subfigure}
    \begin{subfigure}[]{\linewidth}
        \includegraphics[width=\linewidth]{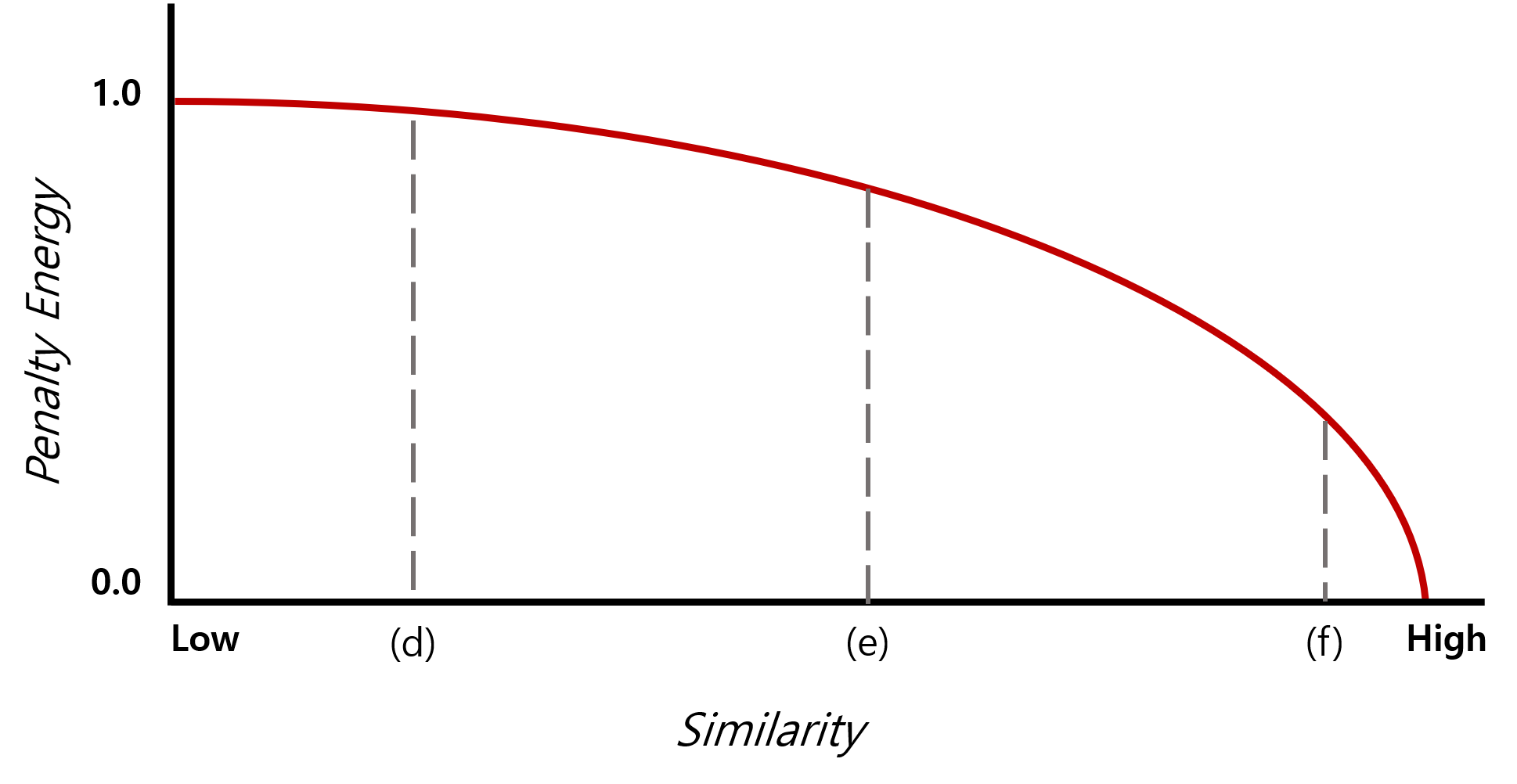}
        \subcaption{Penalty energy graph of similarity}
    \end{subfigure}
    \caption{Example of the utilization of penalty energy to refine contour and distance transformation (DT) features. White color indicates a higher value, and the red line indicates the predicted contour. The colored blobs represent equal low probabilities but are colored differently to distinguish between multiple organs. (a) DT features, (b) inversion of clamped DT features, and (c) contour features; (d), (e), and (f) indicate the cases that may appear during the training; (g) shows the trend of our penalty energy in terms of the similarity, which indicates the contour loss. The case shown in (d) has a high penalty as the contour features are overlapping the background area, which has high values. Similarly, the case illustrated in (f) has a low penalty energy. The penalty energy (\ref{eq:penalty}) performs an important role as the contour feature responses do not appear in the background to identify the exact boundary line.}
    \label{fig:Method_penalty}
\end{figure}

\vspace{2mm}
\noindent \textit{\textbf{Penalty Energy to Refine Features}}
The outputs of the CTN and DTTN may produce noisy contour probability map and DT features, respectively. Because more precise features can build a fine attention map at a shape-aware attention module, we incorporated the penalty energy to regularize the output of the CTN and DTTN. To regularize the feature map responses, we first applied a sigmoid function to the contour features. Thus, each voxel feature represents the contour probability in the range of $[0, 1]$. In the case of DT features, we clamped the DT image to distinguish between the foreground and background voxels. Subsequently, we inverted the DT features by subtractingthe value from 1 to flip the foreground and background. Finally, the penalty energy for regularization is defined by calculating the product of the contour features with the subtracted inverse DT features. The penalty energy is obtained as follows:
\begin{equation}
    E_{p}=\sigma(f_{c}) \otimes (1-clamp(f_{dt}, 0, 1)),
    \label{eq:penalty}
\end{equation}
where the $clamp(f, a, b)$ function maintains only the values of $f$ within the range $[a,b]$; moreover, $f_{c}$ and $f_{dt}$ indicate the contour and DT features, respectively. This penalty energy becomes zero if the responses of the contour features do not appeared in the background area of the DT features. Fig. \ref{fig:Method_penalty} illustrates an example of the penalty energy. Thus, it can reduce the false-positive responses from the contour probabilistic map (i.e., $f_{c}$), primarily by assigning more penalty in the background when compared to the proximate regions.\par

\begin{figure}[t!]
    \centering
    \includegraphics[width=\linewidth]{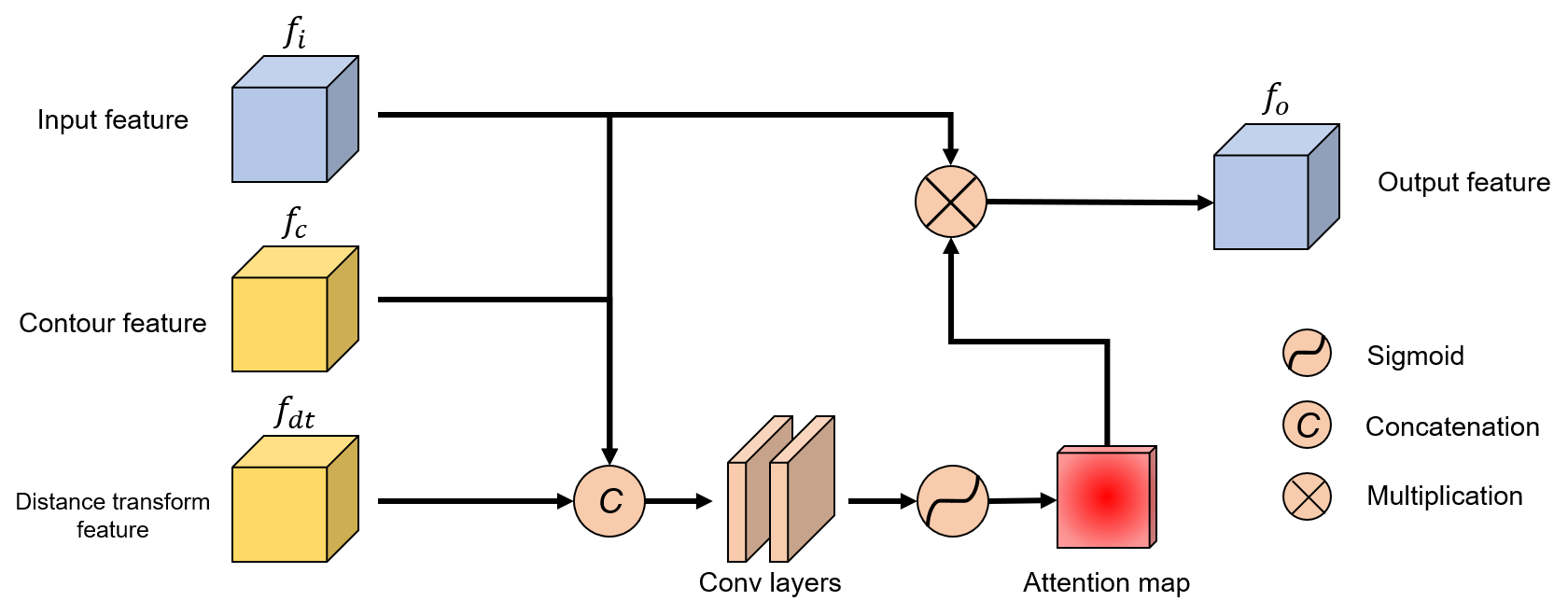}
    \caption{Diagram of shape-aware attention module (Red block in Fig. \ref{fig:Our_arch}). The module generates the attention map from the composition features of DT and contour. The attention map instructs the model to focus on the boundary of the objects.}
    \label{fig:Our_attention}
\end{figure}

\vspace{2mm}
\noindent \textit{\textbf{Shape-aware Attention}}
Finally, the attention mechanism was employed in our CDA-Net. The final shape-aware attention module exploits three different features from the previous layers: 1) contour probabilistic map from CTN, 2) DT feature map from DTTN, and 3) the features from the backbone network. The final attention module can be viewed as \textit{contour- and distance- transform-guided shape-aware attention module}, which facilitates each shape and contour feature. Fig. \ref{fig:Our_attention} illustrates the details of our shape-aware attention block.
The input features (i.e., features from the backbone network) are concatenated with the contour and DT features. Subsequently, the attention map is generated by employing a series of convolutional layers and a sigmoid function. The attention map is finally multiplied by an element-wise operator (i.e., $\otimes$) with the input feature. The output feature response are more representative in the object area primarily owing to the mixed features (i.e., contour and DT). The DT feature guides not only DTTN but also the attention map, which can significantly aid the generation process of the attention map to be more representative inside the organs. We can represent $f_o$ the output of the shape-aware attention as:
\begin{equation}
    f_{o} = f_{i} \otimes A,
    \label{eq:our_att_out}
\end{equation}
where $A$ is the attention map, 
\begin{equation}
    A = \sigma(\mathcal{C}(f_i;f_c;f_{dt})),
    \label{eq:our_att_map}
\end{equation}
where $\sigma(\cdot)$ denotes the nonlinear sigmoid activation function, $\mathcal{C}(\cdot)$ is the convolutional layer, and $(A;B)$ indicates the concatenation of $A$ and $B$.\par

\subsection{Overall Loss Function}
For cardiac segmentation, as mentioned above, the contour map and DT are used to formulate the loss function to predict the precise results from the model. The details of the loss function are described subsequently. \par
First, the data imbalance problem between the foreground and background voxels should be addressed in the segmentation. Therefore, we used Dice loss \cite{milletari2016vnet} to formulate our loss function. That is, to address the multi-class segmentation problem, the generalized DSC (GD) loss \cite{Sudre_2017} is employed. Given the reference image $R$ with voxel values $r_n$, and the predicted probabilistic map $P$ with elements $p_n$, GD loss takes the following form:
\begin{equation}
    L_{GD}=1-2\cfrac{\sum_{c=1}^{N_{cls}}{w_c}\sum_n{r_{cn}p_{cn}}}{\sum_{c=1}^{N_{cls}}{w_c\sum_n{r_{cn}+p_{cn}}}}
    \label{eq:loss_gde}
\end{equation}
where $N_{cls}$ indicates the number of classes to be segmented, and $w_c$ is used to assign weight to each label. $w_c$ is usually determined by the number of class voxels, denoted by $w_c=1/{(\sum_{n=1}^{N}{r_{cn}})^2}$. The GD loss is computed between the final probabilistic map outputs of our model and the ground-truth labels, denoted as $L_O$.\par
Subsequently, to regress the contour probability map and DT feature, BCE loss (\ref{eq:loss_bce}) and MSE loss (\ref{eq:loss_mse}) are applied for the contour loss $L_C$ and the DT loss $L_{DT}$, respectively. The penalty energy term $E_p$ is also combined with the our loss function. \par
Finally, we formulated our loss function to minimize for CDA-Net. We added all the above terms with the weighting coefficients. The final loss function is as follows:
\begin{equation}
    L=\lambda_{1}L_{O}+\lambda_{2}L_{C}+\lambda_{3}L_{DT}+\lambda_{4}E_{p}.
    \label{eq:loss_CDANet}
\end{equation}
In this equation, $\lambda$ is the weight of each term. We used $\lambda_1=1, \lambda_2=20, \lambda_3=10,$ and $\lambda_4=1$ in the experiments.

\subsection{Learning the Network}
To train the CTN and DTTN, the contour images and DTs of the ground-truth labels were precomputed for each substructure. All the subvolumes were concatenated by channel dimension, and formed an $N \times D \times H \times W$ structure, where N is the number of substructures, and D, H, and W are the depth, height, and width, respectively. We used contour images to represent the boundary details, and the foreground DT (FDT) to utilize shape information. To acquire the ground-truth contour image, the Prewitt filter was applied in the a 3D direction to generate a gradient image. The FDT was computed using a linear time algorithm \cite{dt_algo}. \par
To overcome the limitation of GPU memory, we resized the image to $128 \times 128 \times 64$ voxels for the model input. The input images were pre-processed with fixed windowing values ranging between -300 and 1000. We normalized the image voxel values to [0,1]. To generalize the model, Gaussian noise, rotation, and cutout \cite{Shorten2019} were randomly used for data augmentation. \par
We attempted to train our model to minimize the loss $L$ (\ref{eq:loss_CDANet}). While training the network, we used the Adam optimizer \cite{kingma2017adam} with a learning rate of 1e-3. The segmentation outputs were obtained by applying the softmax function to the final feature maps. All training experiments was conducted on an Intel 10 core 19-7900X processor and a 24GB Nvidia Titan RTX GPU machine with 128GB memory. We implemented the proposed network using the PyTorch framework \cite{pytorch2019}. 

\begin{table*}[ht!]
\centering
{
\begin{tabular}{l|c||c|c|c|c|c|c|c|c}
Network & Metric   & LA              & LV              & RA              & RV              & LV-myo          & AA              & PA              & WH               \\ 
\hhline{==::========}
\multirow{2}{*}{U-Net}      & DSC & 0.8951          & 0.8347          & 0.8172          & 0.7712          & 0.7845          & 0.8081          & 0.7111          & 0.8184           \\
& JI & 0.8138          & 0.7170          & 0.6984          & 0.6353          & 0.6508          & 0.7049          & 0.5681          & 0.6955           \\
\hline
\multirow{2}{*}{VoxResNet} & DSC & 0.8750          & 0.7880          & 0.7197          & 0.7328          & 0.7461          & 0.7525          & 0.6062          & 0.7655           \\
& JI & 0.7865          & 0.6559          & 0.5735          & 0.5850          & 0.6009          & 0.6400          & 0.4503          & 0.6246           \\
\hline
\multirow{2}{*}{DenseVNet} & DSC & 0.8708          & 0.7998          & 0.8288          & 0.7602          & 0.7921          & 0.7716          & 0.6810          & 0.8100           \\
& JI & 0.7762          & 0.6764          & 0.7265          & 0.6315          & 0.6629          & 0.6746          & 0.5453          & 0.6866           \\
\hline
\multirow{2}{*}{AGU-Net} & DSC & 0.8575          & 0.7921          & 0.8448          & 0.7950          & 0.8017          & 0.8142          & 0.7180          & 0.8203           \\
& JI   & 0.7693          & 0.6809          & 0.7526          & 0.6775          & 0.6806          & 0.7208          & 0.5979          & 0.7059           \\
\hline
\multirow{2}{*}{Ours} & DSC & \textbf{0.9005} & \textbf{0.8656} & \textbf{0.8935} & \textbf{0.8638} & \textbf{0.8242} & \textbf{0.8736} & \textbf{0.7915} & \textbf{0.8700} \\
& JI      & \textbf{0.8221} & \textbf{0.7709} & \textbf{0.8113} & \textbf{0.7657} & \textbf{0.7118} & \textbf{0.7999} & \textbf{0.6775} & \textbf{0.7733} 
\end{tabular}
}
\caption{Dice similarity coefficient and Jaccard index score of contour- and distance-transform-guided attention network and the state-of-the-art methods on the Multi-Modality Whole Heart Segmentation 2017 (MM-WHS 2017) computed tomography image dataset.}
\label{tab:comp_sota}
\end{table*}

\section{Experimental Results}
\subsection{Datasets}
We trained our model with the Multi-Modality Whole Heart Segmentation 2017 (MM-WHS 2017) dataset\footnote{available on http://www.sdspeople.fudan.edu.cn/zhuangxiahai/0/mmwhs/} \cite{ZHUANG201677, Zhuang1900}, which provides 60 CT and 60 MRI cardiac images. From each set of 60 images, 20 images with label-annotated data were included for training and 40 images were used for testing. We used only the CT dataset for validation. The Whole Haert Segmentation (WHS) ground-truth data were manually labeled by well-trained students majoring in biomedical engineering or medical physics \cite{ZHUANG201677}. Seven substructures were selected as areas of interest in the WHS study, including:\newline
1) LV: the left ventricular cavity\newline
2) RV: the right ventricular cavity\newline
3) LA: the left atrial cavity\newline
4) RA: the right atrial cavity\newline
5) LV-myo: the myocardium of left ventricle\newline
6) AA: the ascending aorta trunk\newline
7) PA: the pulmonary artery trunk\newline

The training dataset was split into a training/validation set to generalize the model using n-fold cross-validation.

\subsection{Evaluation Metrics}
The segmentation results were evaluated using the DSC and Jaccard index (JI). Given the binary labeled masks X and Y, we define the DSC and JI as follows:
\begin{equation}
    DSC(X,Y)=\cfrac{2|X\cap Y|}{|X|+|Y|}
    \label{eq:dice}
\end{equation}
\begin{equation}
    J(X,Y)=\cfrac{|X\cap Y|}{|X\cup Y|}=\cfrac{|X\cap Y|}{|X|+|Y|-|X\cap Y|}
    \label{eq:jaccard}
\end{equation}

We also evaluated the surface distance metrics, i.e., 95\% Hausdorff distance (HD) and average symmetric surface distance (ASSD), to demonstrate the performance of the proposed shape-aware attention module. The 95\% HD was processed without 5\% of the outlying voxels because it is more robust if noisy outliers are avoided. The HD is defined as follows:
\begin{equation}
    HD(X,Y)=\max\{ \max_{{s_X\in S_X}} d(s_X, S_Y) + \max_{s_Y \in S_Y} d(s_Y, S_X) \}
    \label{eq:hd}
\end{equation}
where $d(p, S_X)$ is the shortest distance from an arbitrary voxel $p$ to a set of surfaces $S_X$. 
\begin{equation}
    d(p, S_X)=\min_{s_X \in S_X} ||p-s_X||_2
    \label{eq:short_dist}
\end{equation}
We can also define distance function between two sets as
\begin{equation}
    D(S_X, S_Y)=\sum_{s_X \in S_X}d(s_X, S_Y)
    \label{eq:dist}
\end{equation}
Then, the ASSD can be defined as follows:
\begin{equation}
    ASSD(X, Y)=\cfrac{1}{|S_X|+|S_Y|}(D(S_X, S_Y) +D(S_Y, S_X))
    \label{eq:assd}
\end{equation}
To prove the reduction of false-positive responses, sensitivity and precision were calculated as follows:
\begin{equation}
    S = \cfrac{TP}{TP + FN}
    \label{eq:sensitivity}
\end{equation}
\begin{equation}
    P = \cfrac{TP}{TP + FP}
    \label{eq:precision}
\end{equation}
where TP, FP, and FN are the numbers of true-positive, false-positive, and false-negative output voxels, respectively.\par

\begin{figure}[t]
    \centering
    \includegraphics[width=\linewidth]{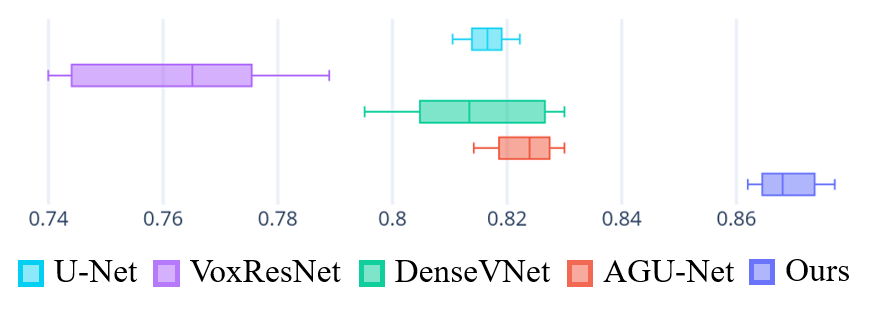}
    \caption{Box plots of dice similarity coefficient score on Whole Heart (WH) segmentation.}
    \label{fig:comp_box_plot}
\end{figure}

\subsection{Comparison}

To assess the cardiac segmentation performance of our model, we compared our proposed CDA-Net with other state-of-the-art networks, i.e., U-Net \cite{3dunet2016}, Voxelwise Residual Network (VoxResNet) \cite{chen2016voxresnet}, DenseVNet \cite{Gibson2018}, and Attention Gate U-Net (AGU-Net) \cite{oktay2018attention}. \par

\vspace{2mm}
\noindent \textit{\textbf{Quantitative Results:}}
Table \ref{tab:comp_sota} lists the quantitative results of cardiac segmentation. The DSC and JI score were computed for all cardiac substructures including LV, RV, LA, RA, LV-myo, AA, PA, and WH. Our proposed network outperformed other state-of-the-art networks in terms of the DSC and JI. In particular, CDA-Net achieved high score on in terms of identifying uneven boundary structures i.e., LV-myo, AA, and PA by finding shape-aware features. However, the shape-prior features and self-attention mechanism showed inferior performance in cardiac multiorgan segmentation. DenseVNet which is trained by shape-prior features using an additional tensor, did not succeed in segmenting accurate cardiac organs because the $12^3$ resolutions of features struggled to learn the details of object shapes. AGU-Net also failed to produce fine segmentation. The primary reason for the failure is that the self-attention mechanism worked to mine common features; consequently, the details of the object (i.e., boundary areas) were easily neglected. The box plots for the paired t-test are illustrated in Fig. \ref{fig:comp_box_plot}. \par

\begin{figure*}
    \begin{center}
    \setlength{\tabcolsep}{4pt}
    \begin{tabular}{ccccccc}
    \includegraphics[width=2.3cm]{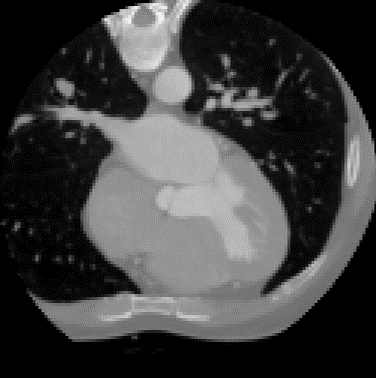} &
    \includegraphics[width=2.3cm]{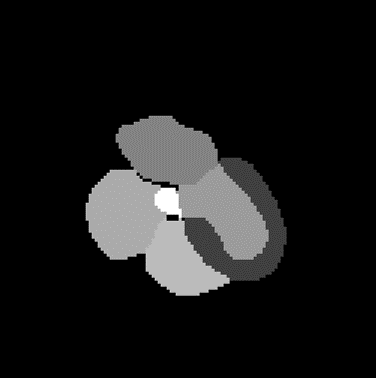} &
    \includegraphics[width=2.3cm]{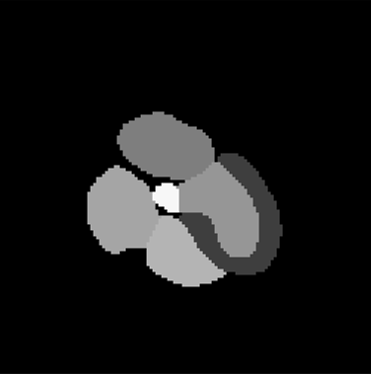} &
    \includegraphics[width=2.3cm]{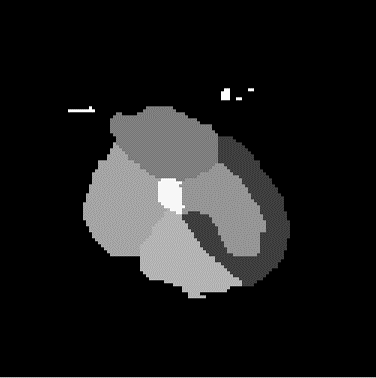} &
    \includegraphics[width=2.3cm]{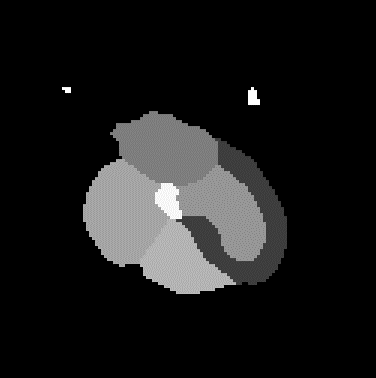} &
    \includegraphics[width=2.3cm]{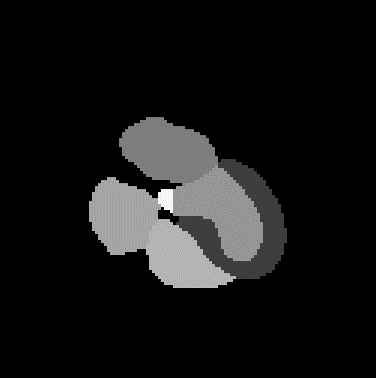} &
    \includegraphics[width=2.3cm]{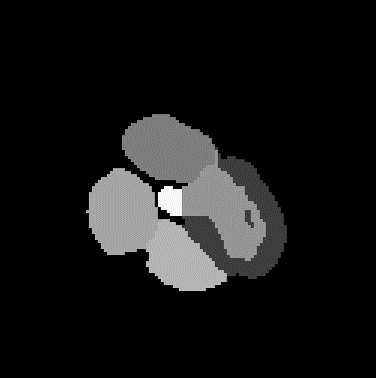} \\
    \includegraphics[width=2.3cm]{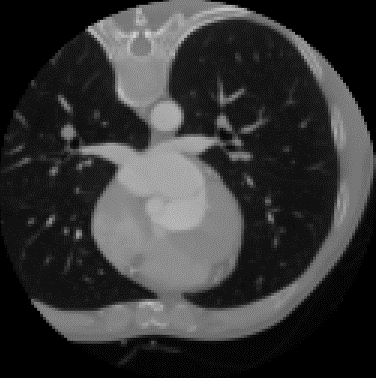} &
    \includegraphics[width=2.3cm]{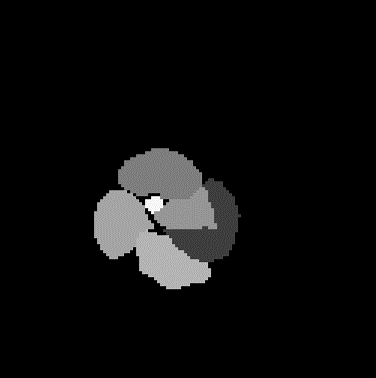} &
    \includegraphics[width=2.3cm]{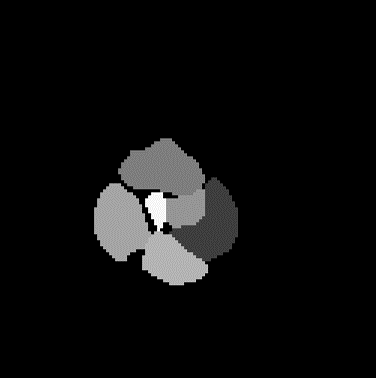} &
    \includegraphics[width=2.3cm]{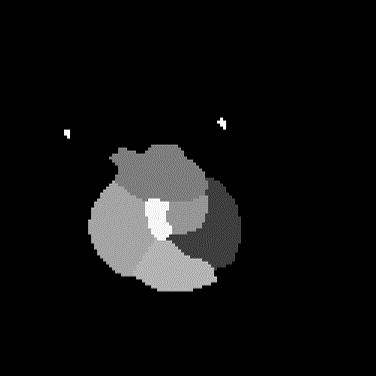} &
    \includegraphics[width=2.3cm]{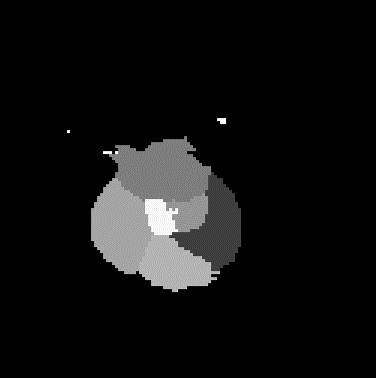} &
    \includegraphics[width=2.3cm]{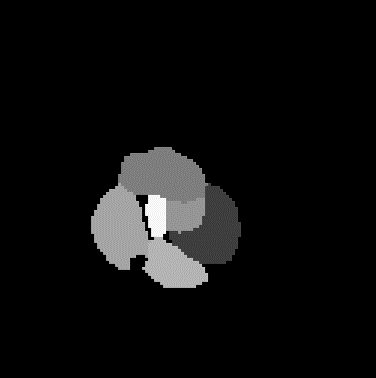} &
    \includegraphics[width=2.3cm]{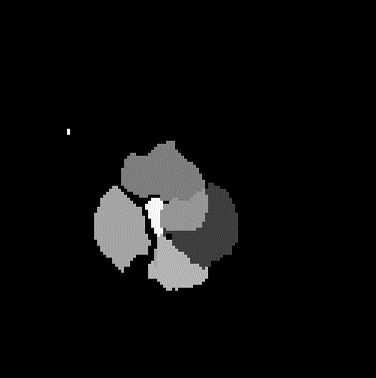} \\
    \includegraphics[width=2.3cm]{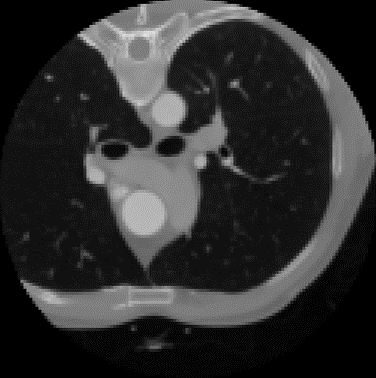} &
    \includegraphics[width=2.3cm]{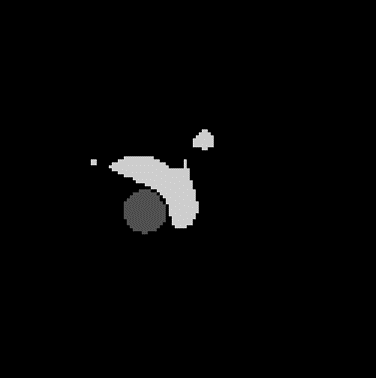} &
    \includegraphics[width=2.3cm]{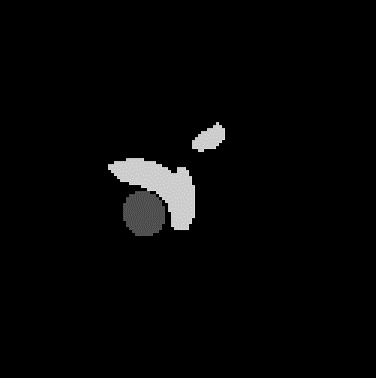} &
    \includegraphics[width=2.3cm]{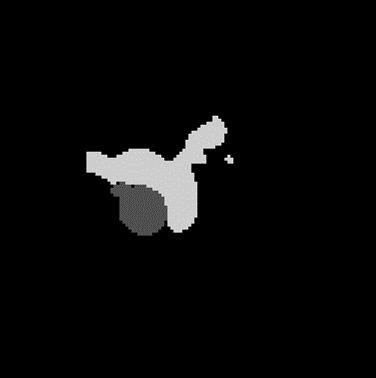} &
    \includegraphics[width=2.3cm]{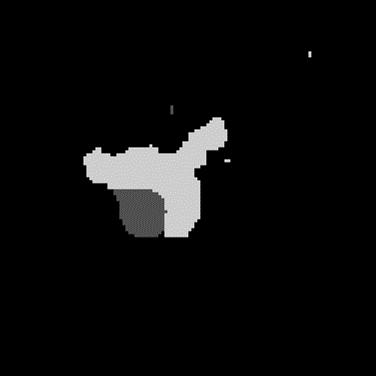} &
    \includegraphics[width=2.3cm]{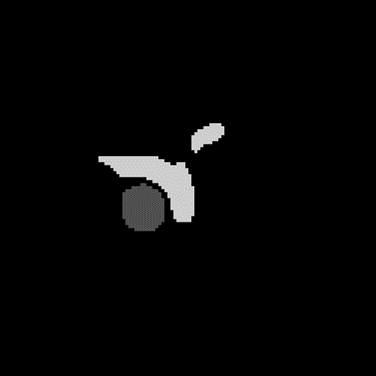} &
    \includegraphics[width=2.3cm]{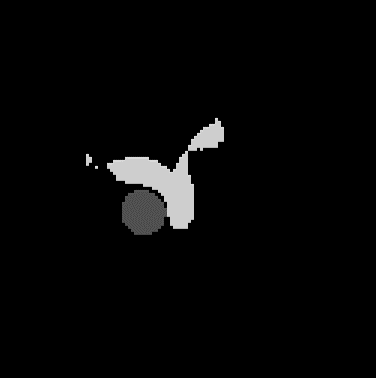} \\
    \makecell{CT\\Image} & \makecell{Ground\\Truth} & Ours & U-Net & VoxResNet & DenseVNet & AGU-Net \\
    \end{tabular}
    \end{center}
    \caption{Axial slices of cardiac structures segmentation results using different networks}
    \label{fig:Exp_vis_2d}
\end{figure*}

\begin{figure*}
    \minipage{0.16\textwidth}
        \begin{subfigure}[]{\linewidth}
            \centering
            \includegraphics[width=2.5cm]{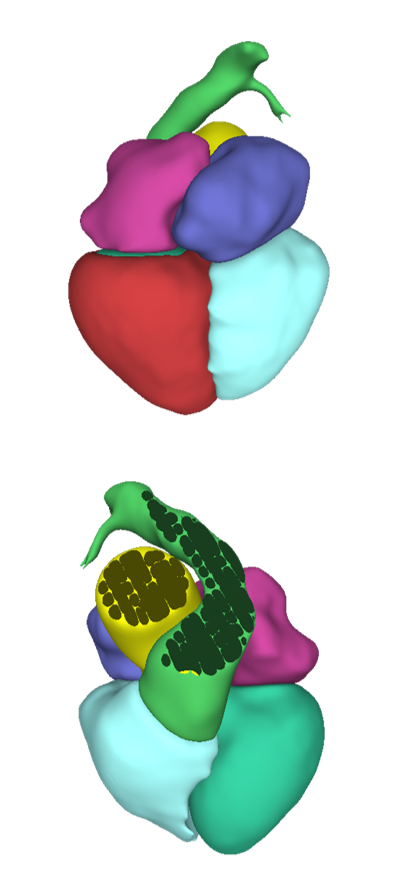}
            \caption{Ground-Truth}
        \end{subfigure}
    \endminipage\hfill
    \minipage{0.16\textwidth}
        \begin{subfigure}[]{\linewidth}
            \centering
            \includegraphics[width=2.5cm]{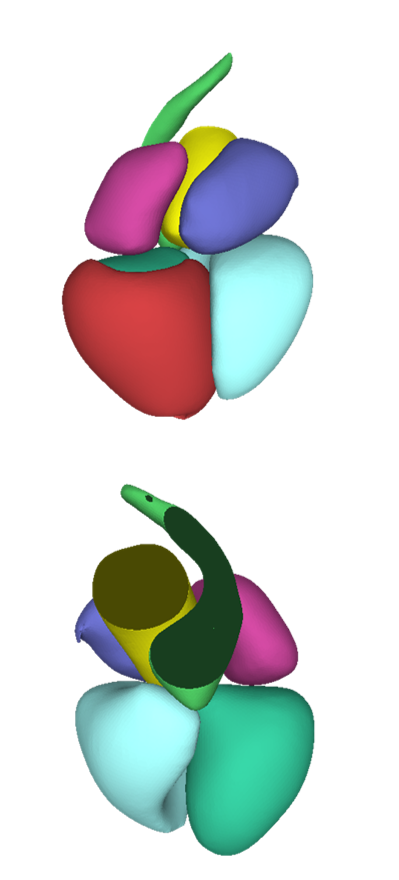}
            \caption{Ours}
        \end{subfigure}
    \endminipage\hfill
    \minipage{0.16\textwidth}
        \begin{subfigure}[]{\linewidth}
            \centering
            \includegraphics[width=2.5cm]{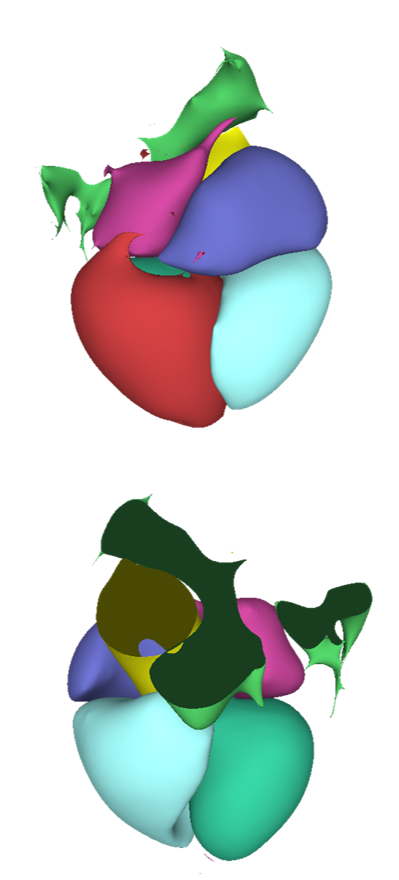}
            \caption{U-Net}
        \end{subfigure}
    \endminipage\hfill
    \minipage{0.16\textwidth}
        \begin{subfigure}[]{\linewidth}
            \centering
            \includegraphics[width=2.5cm]{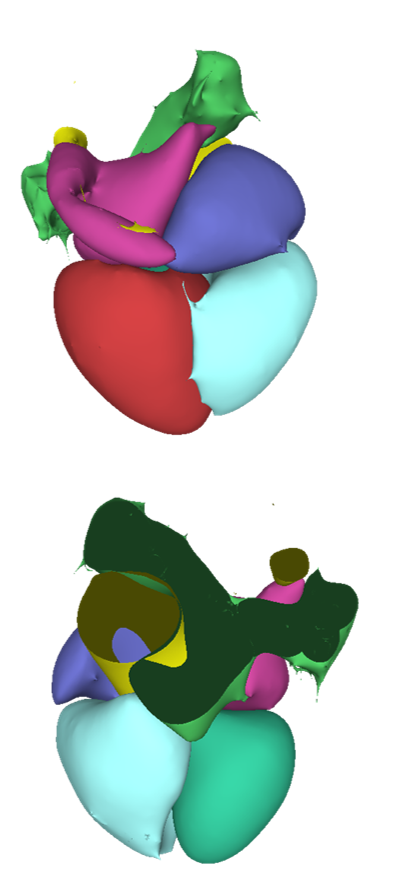}
            \caption{VoxResNet}
        \end{subfigure}
    \endminipage\hfill
    \minipage{0.16\textwidth}
        \begin{subfigure}[]{\linewidth}
            \centering
            \includegraphics[width=2.5cm]{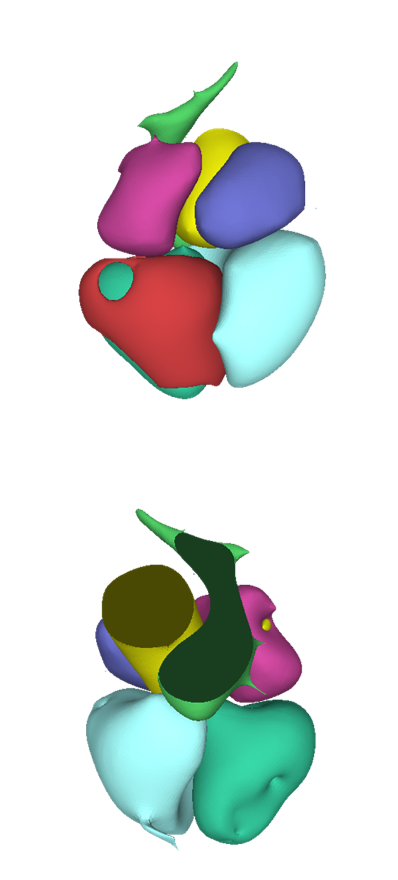}
            \caption{DenseVNet}
        \end{subfigure}
    \endminipage\hfill
    \minipage{0.16\textwidth}
        \begin{subfigure}[]{\linewidth}
            \centering
            \includegraphics[width=2.5cm]{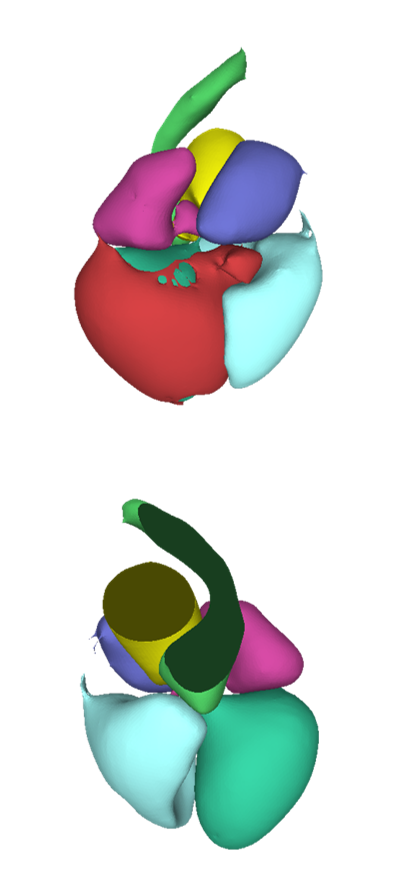}
            \caption{AGU-Net}
        \end{subfigure}
    \endminipage\hfill
    
    \caption{Visualization of the cardiac structures segmentation results obtained using networks. The second row of each image is shown without the myocardium of the left ventricle.}
    \label{fig:Exp_vis}
\end{figure*}

\vspace{2mm}
\noindent \textit{\textbf{Qualitative Results:}}
Figs. \ref{fig:Exp_vis_2d} and \ref{fig:Exp_vis} show the visualization of the cardiac the segmentation results. Fig. \ref{fig:Exp_vis_2d} shows the axial slice of segmentation results and Fig. \ref{fig:Exp_vis} illustrates the volume and surface of the predicted labels.
Notably, U-Net and VoxResNet, which have no shape features, exhibited false-positive responses outside of the heart shapes. Conversely, the networks that learned the heart shapes produced less noisy and smooth surface results. Although, DenseVNet and AGU-Net reduced the false-positive responses, they still had difficulty in determining the boundaries among the substructures. When compared to other results, our proposed network detected precise boundaries without any false-positive responses. The results indicate that the shape-aware attention module successfully suppressed the mis-segmented outputs and concentrated on the boundary area of the target object.\par

\subsection{Ablation Study}
We appended ablation studies to evaluate the proposed method. Table \ref{tab:comp_ab} lists the comparison results of the ablations. \textit{base} indicates a U-Net backbone network with an appended V-transition module. In \textit{base+CBAM}, the shape-aware attention module is replaced with a CBAM\cite{cbam}. In addition, ablation experiments were conducted by excluding certain of modules in CDA-Net. \textit{base+CTN} is the applied attention mechanism with only the contour probability map. Conversely, \textit{base+DTTN} is composed of the backbone and DTTN for the DT feature. \textit{base+CTN+DTTN} and \textit{base+CTN+DTTN+penalty} were compared to verify the effectiveness of the proposed penalty function.\par
As listed in Table \ref{tab:comp_ab}, our model outperforms the other ablations. \textit{base+CTN} and \textit{base+DTTN} showed minimal improvement in the DSC score when compared to the base network. However, \textit{base+CTN+DTTN} showed a significant improvement when compared to the base network. In particular, our model with the contour probability map and DT feature achieved high scores in 95\% HD, ASSD, and precision, indicating that our proposed shape-aware attention module with CTN and DTTN significantly reduced false-positive responses. \par
\begin{figure*}
    \begin{center}
    \setlength{\tabcolsep}{4pt}
    \begin{tabular}{ccccccc}
    \includegraphics[width=2.3cm]{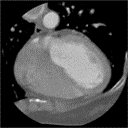} &
    \includegraphics[width=2.3cm]{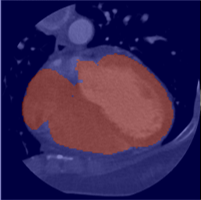} &
    \includegraphics[width=2.3cm]{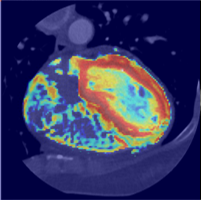} &
    \includegraphics[width=2.3cm]{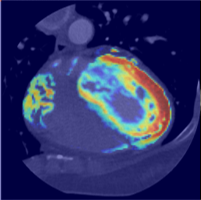} &
    \includegraphics[width=2.3cm]{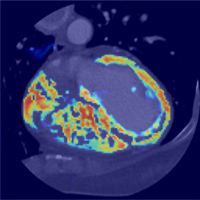} &
    \includegraphics[width=2.3cm]{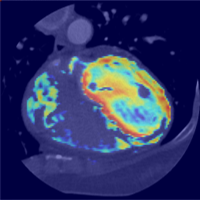} &
    \includegraphics[width=2.3cm]{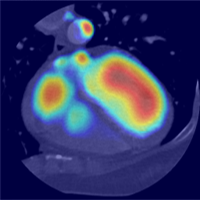} \\
    \includegraphics[width=2.3cm]{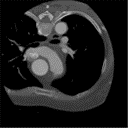} &
    \includegraphics[width=2.3cm]{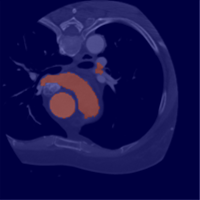} &
    \includegraphics[width=2.3cm]{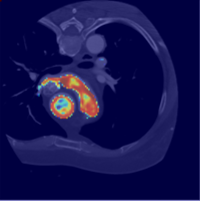} &
    \includegraphics[width=2.3cm]{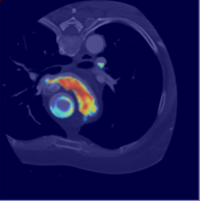} &
    \includegraphics[width=2.3cm]{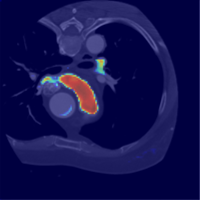} &
    \includegraphics[width=2.3cm]{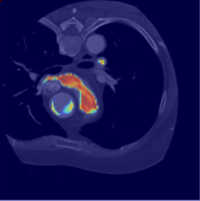} &
    \includegraphics[width=2.3cm]{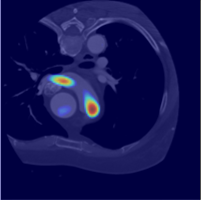} \\
    \makecell{CT\\Image} & \makecell{Ground\\Truth} & Ours & \makecell{base\\+CTN+DTTN} & \makecell{base\\+DTTN} & \makecell{base\\+CTN} & \makecell{base\\+CBAM} \\
    \end{tabular}
    \end{center}
    \caption{Visualization of attention maps from CDA-Net and its ablations}
    \label{fig:ab_heatmap}
\end{figure*}

Figure \ref{fig:ab_heatmap} shows the attention map of our network and its ablations. The traditional self-attention mechanism CBAM shows sufficient performance while identifying the blob of the target object but reveals weaknesses in detecting the precise boundary of the object. Self-attention with only the contour probability map or DT feature weakens the shape-prior information and the details of the boundary contexts. When we employed both auxiliary modules, the attention map showed a stronger response in the boundary area. Moreover, our proposed penalty energy enhanced the contour responses and preserved the shape-prior information. Accordingly, the contour probability map and DT mutually affected the construction of a complementary feature that forms a fine attention map that has strong responses on the boundary area. In addition, the proposed penalty energy could strengthen the attention map.\par

\begin{table*}
\centering
{
    \begin{tabular}{l|c||c|c|c|c|c|c|c|c}
    Network & Metric & LA & LV & RA & RV & LV-myo & AA & PA & WH \\ 
    \hhline{==::========}
    \multirow{5}{*}{base} & DSC & 0.867 & 0.889 & 0.816 & 0.834 & 0.805 & 0.938 & 0.814 & 0.910 \\
     & HD95 & 2.222 & 3.747 & 3.244 & 5.610 & 4.828 & 3.990 & 6.676 & 3.401 \\
     & ASSD & 0.855 & 1.083 & 0.950 & 1.481 & 1.461 & 0.829 & 1.473 & 0.982 \\
     & Sensitivity & 0.836 & 0.910 & 0.909 & 0.883 & 0.880 & 0.950 & 0.865 & 0.952 \\
     & Precision & 0.790 & 0.841 & 0.864 & 0.781 & 0.808 & 0.934 & 0.782 & 0.879 \\
    \hline
    \multirow{5}{*}{base+CBAM \cite{cbam}} & DSC & 0.870 & 0.902 & 0.810 & 0.824  & 0.809 & 0.923 & 0.792 & 0.910 \\
    & HD95 & 2.760 & 4.669  & 5.130 & 5.195 & 5.740 & 4.746  & 9.162 & 4.278  \\
    & ASSD & 0.977 & 1.193 & 0.971 & 1.551 & 1.701 & 0.944 & 1.840 & 1.086 \\
    & Sensitivity & 0.837  & 0.913 & 0.905 & 0.870 & 0.903 & 0.918 & 0.861 & 0.952 \\
    & Precision & 0.795 & 0.836 & 0.906 & 0.775 & 0.769 & 0.933 & 0.749 & 0.874 \\
    \hline
    \multirow{5}{*}{base+CTN} & DSC & 0.878 & 0.910 & 0.826 & 0.852 & 0.826 & 0.941 & 0.817 & 0.923 \\
    & HD95 & 2.211 & 3.625 & 2.329 & 5.277 & 6.283 & 2.999 & \textbf{6.570} & 3.432 \\
    & ASSD  & 0.844 & 1.090 & 0.824 & 1.452 & 1.542 & 0.656 & 1.365 & 0.930  \\
    & Sensitivity & 0.844 & 0.915 & 0.916 & 0.876 & 0.844 & \textbf{0.944} & 0.853 & 0.948 \\
    & Precision & 0.810 & 0.840 & 0.903 & 0.781 & 0.832 & 0.939 & 0.767 & 0.892 \\
    \hline
    \multirow{5}{*}{base+DTTN} & DSC & 0.868 & 0.907 & 0.823 & 0.859 & 0.822 & 0.939 & 0.815 & 0.924 \\
    & HD95 & 2.687 & 3.991 & 2.853 & 6.163 & 4.331 & 3.412 & 7.274 & 3.168 \\
    & ASSD & 0.877 & 1.125 & 0.910 & 1.568 & 1.286 & 0.721 & 1.378 & 0.923 \\
    & Sensitivity & 0.846 & 0.906 & 0.902 & 0.863 & 0.896 & 0.925 & 0.847 & 0.944 \\
    & Precision & 0.816	& 0.835 & 0.923 & 0.798 & 0.843 & 0.957 & 0.786 & 0.904 \\
    \hline
    \multirow{5}{*}{base+CTN+DTTN} & DSC & 0.872 & 0.912 & \textbf{0.839} & 0.862 & \textbf{0.830} & 0.951 & 0.821 & 0.927 \\
    & HD95 & 2.126 & 3.863 & 2.336 & 5.217 & \textbf{4.089} & 2.297 & 6.792 & 3.013 \\
    & ASSD & 0.844 & 1.103 & 0.947 & 1.424 & \textbf{1.235} & 0.585 & 1.341 & 0.882 \\
    & Sensitivity & 0.845 & 0.933 & 0.917 & \textbf{0.877} & \textbf{0.907} & \textbf{0.944} & 0.870 & \textbf{0.958} \\
    & Precision & \textbf{0.831} & 0.840 & \textbf{0.916} & 0.818 & 0.842 & 0.967 &	0.796 &	0.909  \\
    \hline
    \multirow{5}{*}{\makecell{base+CTN+DTTN+penalty\\(Ours)}} & DSC & \textbf{0.888} & \textbf{0.917} & \textbf{0.836} & \textbf{0.869} & \textbf{0.831} & 0.946 & \textbf{0.834} & \textbf{0.935} \\
    & HD95 & \textbf{2.063} & \textbf{3.218} & 1.989 & \textbf{3.959} & 4.410 & \textbf{2.000} & \textbf{6.608} & \textbf{2.659} \\
    & ASSD & \textbf{0.793} & \textbf{0.946} & \textbf{0.750} & \textbf{1.287} & 1.267 & \textbf{0.544} & \textbf{1.258} & \textbf{0.810} \\
    & Sensitivity & \textbf{0.857} & 0.928 & \textbf{0.928} & 0.872 & 0.890 & 0.938 & \textbf{0.875} & \textbf{0.957} \\
    & Precision & 0.820 &	\textbf{0.861} &	\textbf{0.912} &	\textbf{0.824} &	\textbf{0.857} & \textbf{0.968} & \textbf{0.821} & \textbf{0.915} \\
    \end{tabular}
}
\caption{Assessment of ablation networks based on ground-truth}
\label{tab:comp_ab}
\end{table*}

\begin{figure}[htb]
    \centering
    \begin{subfigure}[]{\linewidth}
        \centering
        \includegraphics[width=\linewidth]{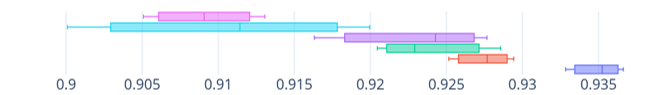}
        \caption{DSC}
    \end{subfigure}
    \begin{subfigure}[]{\linewidth}
        \centering
        \includegraphics[width=\linewidth]{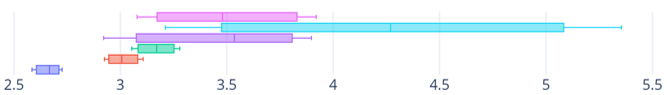}
        \caption{HD95}
    \end{subfigure}
    \begin{subfigure}[]{\linewidth}
        \centering
        \includegraphics[width=\linewidth]{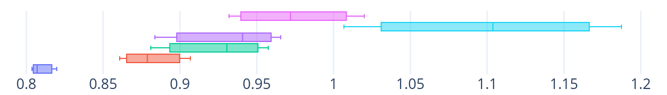}
        \caption{ASSD}
    \end{subfigure}
    \begin{subfigure}[]{\linewidth}
        \centering
        \includegraphics[width=\linewidth]{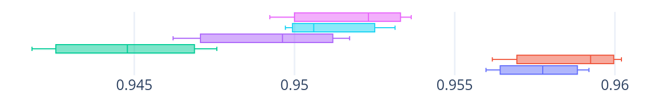}
        \caption{Sensitivity}
    \end{subfigure}
    \begin{subfigure}[]{\linewidth}
        \centering
        \includegraphics[width=\linewidth]{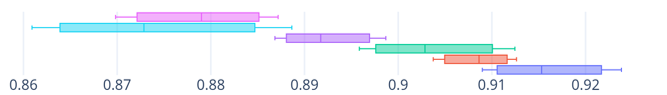}
        \caption{Precision}
    \end{subfigure}
    \begin{subfigure}[]{\linewidth}
        \centering
        \includegraphics[width=0.7\linewidth]{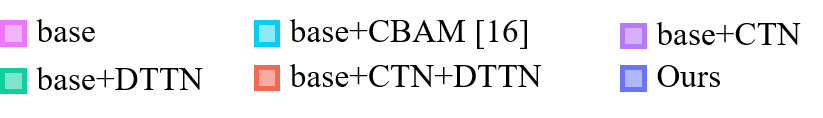}
    \end{subfigure}
    \caption{Box plots of ablation study on WH Segmentation}
    \label{fig:comp_box_plot_b}
\end{figure}

\section{Discussion and Conclusion}
Cardiac segmentation is a challenging task because of the ambiguous multiorgan boundaries. In this study, we focused on resolving ambiguous boundaries to improve cardiac segmentation. The proposed network applies the attention mechanism to utilize the details of boundaries between substructures. Our proposed method attempted to obtain shape-prior features using the attention mechanism because the shape features of each organ mutually complement the accuracy of cardiac segmentation. However, the traditional shape-prior methods are inferior because of the following reasons. DenseVNet \cite{Gibson2018} attached the volume tensor for shape-prior features but showed inferior performance in delineating the boundaries of cardiac organs. Self-attention mechanisms, including CBAM \cite{cbam} and AGU-Net \cite{oktay2018attention}, were also proposed to obtain precise segmentation, but experienced difficulty in determining the exact contour features, even though the self-attention mechanisms are suitable for identifying the blob of an object. Moreover, the shape-prior methods are data-driven algorithms, which indicate that they require various training data distributions for generalized performance. In summary, the preceding networks typically failed to capture the details of the targeted objects, which indicates that it is difficult to achieve high performance on the adjacent multi-organ segmentation task. To complement the typical shape-prior methods, our shape-aware contour attention mechanism attached the contour and DT features to the attention module. We entangled the boundary-aware and shape-prior features using CTN and DTTN, supervised by the contour and DT of the organs, respectively. The DT feature aided the model to easily focus on the blob of the target object, and the contour feature derived the attention map to obtain a high response in the boundary regions. Moreover, penalty energy was applied to enhance the accuracy of the DT and contour features in the complementary relations. Therefore, the proposed shape-aware contour attention method successfully produced a fine attention map that focused on the boundary of the object while maintaining shape-prior information.\par

Additionally, our proposed network significantly reduced the false-positive responses. Conventional deep learning networks \cite{ronneberger2015unet, 3dunet2016, chen2016voxresnet, milletari2016vnet} are based on classifying each voxel, , which is the foundation of future research; however, they failed to reduce the outliers in the background regions. Moreover, when compared to U-Net \cite{ronneberger2015unet} and VoxResNet \cite{chen2016voxresnet}, CDA-Net showed superior performance in reducing outliers and refining the segmentation results, especially in the boundary area. This is because our proposed network \textit{CDA-Net} focused on generating a fine attention map with a has low probability in the background area. Therefore, it can lead to superior performance on multi-organ segmentation problem that require the identification of the exact boundary between organs.\par

In conclusion, the shape features, including the contour and DT of the target objects, minimized the false-positive errors in the segmentation problem. Prior research synthesized shape-prior information and deep features to clarify the extrapolation of the shape features in medical image segmentation. They succeeded in improving segmentation accuracy; however, they failed to detect clear boundaries and the surface of the target organs. From this aspect, a shape-aware contour-guided attention was proposed in this study to refine multiorgan segmentation based on shape features. The proposed shape-aware contour attention method performs an important role in determining the precise boundary, whereas the shape-prior attention only helps in regressing the shape blob approximation. In addition, the proposed method improved the generalization performance when compared to the other data-driven shape-prior-based methods. The contour and DT guided the network to be trained such that the target organs were better delineated, primarily because a precise attention map was obtained in the contour area. Based on the experimental results, the proposed method, which synthesized the contour and DT, led to improvements in terms of robustness and accuracy in medical image segmentation.\par

% \appendices
% \section{Losses during whole training process}
% classification, liver, and tumor segmentation performance acc. plot.

% \section*{Acknowledgment}
% The authors would like to thank Dr. Jin Wook Chung for consistent support.

\ifCLASSOPTIONcaptionsoff
  \newpage
\fi

\bibliographystyle{IEEEtran}
\bibliography{MyBiB}

\begin{thebibliography}{10}
\providecommand{\url}[1]{#1}
\csname url@samestyle\endcsname
\providecommand{\newblock}{\relax}
\providecommand{\bibinfo}[2]{#2}
\providecommand{\BIBentrySTDinterwordspacing}{\spaceskip=0pt\relax}
\providecommand{\BIBentryALTinterwordstretchfactor}{4}
\providecommand{\BIBentryALTinterwordspacing}{\spaceskip=\fontdimen2\font plus
\BIBentryALTinterwordstretchfactor\fontdimen3\font minus
  \fontdimen4\font\relax}
\providecommand{\BIBforeignlanguage}[2]{{%
\expandafter\ifx\csname l@#1\endcsname\relax
\typeout{** WARNING: IEEEtran.bst: No hyphenation pattern has been}%
\typeout{** loaded for the language `#1'. Using the pattern for}%
\typeout{** the default language instead.}%
\else
\language=\csname l@#1\endcsname
\fi
#2}}
\providecommand{\BIBdecl}{\relax}
\BIBdecl

\bibitem{McNamara2019}
K.~Mc~Namara, H.~Alzubaidi, and J.~K. Jackson,
  ``\BIBforeignlanguage{eng}{Cardiovascular disease as a leading cause of
  death: how are pharmacists getting involved?}''
  \emph{\BIBforeignlanguage{eng}{Integrated Pharmacy Research {\&} Practice}},
  vol.~8, pp. 1--11, Feb 2019.

\bibitem{doi:10.1161/CIR.0000000000000757}
S.~S. Virani \emph{et~al.}, ``Heart disease and stroke statistics-2020 update:
  A report from the american heart association,'' \emph{Circulation}, vol. 141,
  no.~9, pp. e139--e596, 2020.

\bibitem{doi:10.1161/CIRCULATIONAHA.107.717033}
H.~C. McGill, C.~A. McMahan, and S.~S. Gidding, ``Preventing heart disease in
  the 21st century,'' \emph{Circulation}, vol. 117, no.~9, pp. 1216--1227,
  2008.

\bibitem{ODONNELL2016761}
M.~J. O'Donnell \emph{et~al.}, ``Global and regional effects of potentially
  modifiable risk factors associated with acute stroke in 32 countries
  (interstroke): a case-control study,'' \emph{The Lancet}, vol. 388, no.
  10046, pp. 761 -- 775, 2016.

\bibitem{simonyan2015deep}
K.~Simonyan and A.~Zisserman, ``Very deep convolutional networks for
  large-scale image recognition,'' 2015.

\bibitem{he2015deep}
K.~He, X.~Zhang, S.~Ren, and J.~Sun, ``Deep residual learning for image
  recognition,'' 2015.

\bibitem{Huang2017}
G.~Huang, Z.~Liu, L.~Van Der~Maaten, and K.~Q. Weinberger, ``Densely connected
  convolutional networks,'' in \emph{2017 IEEE Conference on Computer Vision
  and Pattern Recognition (CVPR)}, 2017, pp. 2261--2269.

\bibitem{ren2016faster}
S.~Ren, K.~He, R.~Girshick, and J.~Sun, ``Faster r-cnn: Towards real-time
  object detection with region proposal networks,'' 2016.

\bibitem{redmon2018yolov3}
J.~Redmon and A.~Farhadi, ``Yolov3: An incremental improvement,'' 2018.

\bibitem{long2015fully}
J.~Long, E.~Shelhamer, and T.~Darrell, ``Fully convolutional networks for
  semantic segmentation,'' 2015.

\bibitem{ronneberger2015unet}
O.~Ronneberger, P.~Fischer, and T.~Brox, ``U-net: Convolutional networks for
  biomedical image segmentation,'' 2015.

\bibitem{milletari2016vnet}
F.~Milletari, N.~Navab, and S.-A. Ahmadi, ``V-net: Fully convolutional neural
  networks for volumetric medical image segmentation,'' 2016.

\bibitem{Gibson2018}
E.~Gibson, F.~Giganti, Y.~Hu, E.~Bonmati, S.~Bandula, K.~Gurusamy, B.~Davidson,
  S.~P. Pereira, M.~J. Clarkson, and D.~C. Barratt,
  ``\BIBforeignlanguage{eng}{Automatic multi-organ segmentation on abdominal ct
  with dense v-networks},'' \emph{\BIBforeignlanguage{eng}{IEEE Transactions on
  Medical Imaging}}, vol.~37, no.~8, pp. 1822--1834, Aug 2018.

\bibitem{oktay2018attention}
O.~Oktay, J.~Schlemper, L.~L. Folgoc, M.~Lee, M.~Heinrich, K.~Misawa, K.~Mori,
  S.~McDonagh, N.~Y. Hammerla, B.~Kainz, B.~Glocker, and D.~Rueckert,
  ``Attention u-net: Learning where to look for the pancreas,'' 2018.

\bibitem{chung2020liver}
M.~Chung, J.~Lee, J.~Lee, and Y.-G. Shin, ``Liver segmentation in abdominal ct
  images via auto-context neural network and self-supervised contour
  attention,'' 2020.

\bibitem{Li2017}
J.~Li, R.~Zhang, L.~Shi, and D.~Wang, ``Automatic whole-heart segmentation in
  congenital heart disease using deeply-supervised 3d fcn,'' in
  \emph{Reconstruction, Segmentation, and Analysis of Medical Images}, M.~A.
  Zuluaga, K.~Bhatia, B.~Kainz, M.~H. Moghari, and D.~F. Pace, Eds.\hskip 1em
  plus 0.5em minus 0.4em\relax Cham: Springer International Publishing, 2017,
  pp. 111--118.

\bibitem{Yu2017}
L.~Yu, J.-Z. Cheng, Q.~Dou, X.~Yang, H.~Chen, J.~Qin, and P.-A. Heng,
  ``Automatic 3d cardiovascular mr segmentation with densely-connected
  volumetric convnets,'' in \emph{Medical Image Computing and Computer-Assisted
  Intervention - MICCAI 2017}, M.~Descoteaux, L.~Maier-Hein, A.~Franz,
  P.~Jannin, D.~L. Collins, and S.~Duchesne, Eds.\hskip 1em plus 0.5em minus
  0.4em\relax Cham: Springer International Publishing, 2017, pp. 287--295.

\bibitem{Zheng2018}
Q.~Zheng, H.~Delingette, N.~Duchateau, and N.~Ayache, ``3-d consistent and
  robust segmentation of cardiac images by deep learning with spatial
  propagation,'' \emph{IEEE Transactions on Medical Imaging}, vol.~37, no.~9,
  pp. 2137--2148, 2018.

\bibitem{KHENED201921}
M.~Khened, V.~A. Kollerathu, and G.~Krishnamurthi, ``Fully convolutional
  multi-scale residual densenets for cardiac segmentation and automated cardiac
  diagnosis using ensemble of classifiers,'' \emph{Medical Image Analysis},
  vol.~51, pp. 21--45, 2019.

\bibitem{Du2019}
X.~Du, S.~Yin, R.~Tang, Y.~Zhang, and S.~Li, ``Cardiac-deepied: Automatic
  pixel-level deep segmentation for cardiac bi-ventricle using improved
  end-to-end encoder-decoder network,'' \emph{IEEE Journal of Translational
  Engineering in Health and Medicine}, vol.~7, pp. 1--10, 2019.

\bibitem{ba2015multiple}
J.~Ba, V.~Mnih, and K.~Kavukcuoglu, ``Multiple object recognition with visual
  attention,'' 2015.

\bibitem{zhou2015learning}
B.~Zhou, A.~Khosla, A.~Lapedriza, A.~Oliva, and A.~Torralba, ``Learning deep
  features for discriminative localization,'' 2015.

\bibitem{chen2016attention}
L.-C. Chen, Y.~Yang, J.~Wang, W.~Xu, and A.~L. Yuille, ``Attention to scale:
  Scale-aware semantic image segmentation,'' 2016.

\bibitem{wang2017residual}
F.~Wang, M.~Jiang, C.~Qian, S.~Yang, C.~Li, H.~Zhang, X.~Wang, and X.~Tang,
  ``Residual attention network for image classification,'' 2017.

\bibitem{pmlr-v37-xuc15}
K.~Xu, J.~Ba, R.~Kiros, K.~Cho, A.~Courville, R.~Salakhudinov, R.~Zemel, and
  Y.~Bengio, ``Show, attend and tell: Neural image caption generation with
  visual attention,'' in \emph{Proceedings of the 32nd International Conference
  on Machine Learning}, ser. Proceedings of Machine Learning Research, F.~Bach
  and D.~Blei, Eds., vol.~37.\hskip 1em plus 0.5em minus 0.4em\relax Lille,
  France: PMLR, 07--09 Jul 2015, pp. 2048--2057.

\bibitem{hu2019squeezeandexcitation}
J.~Hu, L.~Shen, S.~Albanie, G.~Sun, and E.~Wu, ``Squeeze-and-excitation
  networks,'' 2019.

\bibitem{bam}
J.~Park, S.~Woo, J.-Y. Lee, and I.~Kweon, ``Bam: Bottleneck attention module,''
  07 2018.

\bibitem{cbam}
S.~Woo, J.~Park, J.-Y. Lee, and I.~S. Kweon, ``Cbam: Convolutional block
  attention module,'' in \emph{Computer Vision -- ECCV 2018}, V.~Ferrari,
  M.~Hebert, C.~Sminchisescu, and Y.~Weiss, Eds.\hskip 1em plus 0.5em minus
  0.4em\relax Cham: Springer International Publishing, 2018, pp. 3--19.

\bibitem{8489930}
Y.~{Zhuge}, G.~{Yang}, P.~{Zhang}, and H.~{Lu}, ``Boundary-guided feature
  aggregation network for salient object detection,'' \emph{IEEE Signal
  Processing Letters}, vol.~25, no.~12, pp. 1800--1804, 2018.

\bibitem{BORGEFORS1986344}
G.~Borgefors, ``Distance transformations in digital images,'' \emph{Computer
  Vision, Graphics, and Image Processing}, vol.~34, no.~3, pp. 344 -- 371,
  1986.

\bibitem{audebert2019distance}
N.~Audebert, A.~Boulch, B.~L. Saux, and S.~Lefèvre, ``Distance transform
  regression for spatially-aware deep semantic segmentation,'' 2019.

\bibitem{doi:10.1146/annurev-bioeng-071516-044442}
D.~Shen, G.~Wu, and H.-I. Suk, ``Deep learning in medical image analysis,''
  \emph{Annual Review of Biomedical Engineering}, vol.~19, no.~1, pp. 221--248,
  2017.

\bibitem{CHUNG2020103720}
M.~Chung, M.~Lee, J.~Hong, S.~Park, J.~Lee, J.~Lee, I.-H. Yang, J.~Lee, and
  Y.-G. Shin, ``Pose-aware instance segmentation framework from cone beam ct
  images for tooth segmentation,'' \emph{Computers in Biology and Medicine},
  vol. 120, p. 103720, 2020.

\bibitem{10.1007/978-3-030-32692-0_71}
F.~Navarro, S.~Shit, I.~Ezhov, J.~Paetzold, A.~Gafita, J.~C. Peeken, S.~E.
  Combs, and B.~H. Menze, ``Shape-aware complementary-task learning for
  multi-organ segmentation,'' in \emph{Machine Learning in Medical Imaging},
  H.-I. Suk, M.~Liu, P.~Yan, and C.~Lian, Eds.\hskip 1em plus 0.5em minus
  0.4em\relax Cham: Springer International Publishing, 2019, pp. 620--627.

\bibitem{Dangi2019}
S.~Dangi, C.~A. Linte, and Z.~Yaniv, ``\BIBforeignlanguage{eng}{A distance map
  regularized cnn for cardiac cine mr image segmentation},''
  \emph{\BIBforeignlanguage{eng}{Medical Physics}}, vol.~46, no.~12, pp.
  5637--5651, Dec 2019.

\bibitem{3dunet2016}
Özgün Çiçek, A.~Abdulkadir, S.~S. Lienkamp, T.~Brox, and O.~Ronneberger,
  ``3d u-net: Learning dense volumetric segmentation from sparse annotation,''
  2016.

\bibitem{Sudre_2017}
C.~H. Sudre, W.~Li, T.~Vercauteren, S.~Ourselin, and M.~Jorge~Cardoso,
  ``Generalised dice overlap as a deep learning loss function for highly
  unbalanced segmentations,'' \emph{Lecture Notes in Computer Science}, p.
  240–248, 2017.

\bibitem{dt_algo}
C.~R. {Maurer}, {Rensheng Qi}, and V.~{Raghavan}, ``A linear time algorithm for
  computing exact euclidean distance transforms of binary images in arbitrary
  dimensions,'' \emph{IEEE Transactions on Pattern Analysis and Machine
  Intelligence}, vol.~25, no.~2, pp. 265--270, 2003.

\bibitem{Shorten2019}
C.~Shorten and T.~M. Khoshgoftaar, ``A survey on image data augmentation for
  deep learning,'' \emph{Journal of Big Data}, vol.~6, no.~1, p.~60, Jul 2019.

\bibitem{kingma2017adam}
D.~P. Kingma and J.~Ba, ``Adam: A method for stochastic optimization,'' 2017.

\bibitem{pytorch2019}
Paszke \emph{et~al.}, ``Pytorch: An imperative style, high-performance deep
  learning library,'' H.~Wallach, H.~Larochelle, A.~Beygelzimer,
  F.~d\textquotesingle Alch\'{e}-Buc, E.~Fox, and R.~Garnett, Eds.\hskip 1em
  plus 0.5em minus 0.4em\relax Curran Associates, Inc., 2019, pp. 8024--8035.

\bibitem{ZHUANG201677}
X.~Zhuang and J.~Shen, ``Multi-scale patch and multi-modality atlases for whole
  heart segmentation of mri,'' \emph{Medical Image Analysis}, vol.~31, pp. 77
  -- 87, 2016.

\bibitem{Zhuang1900}
X.~Zhuang, ``Challenges and methodologies of fully automatic whole heart
  segmentation: A review,'' \emph{Journal of Healthcare Engineering}, vol.~4,
  p. 981729, Jan 1900.

\bibitem{chen2016voxresnet}
H.~Chen, Q.~Dou, L.~Yu, and P.-A. Heng, ``Voxresnet: Deep voxelwise residual
  networks for volumetric brain segmentation,'' 2016.

\end{thebibliography}

\end{document}